\documentclass[twocolumn,trackchanges]{aastex7}
\accepted{ApJS}

\usepackage{natbib}
\usepackage{amsmath}
\usepackage{multirow}
\usepackage[normalem]{ulem}
\usepackage{xcolor}
\usepackage{xspace}
\usepackage{enumitem}   
\usepackage{graphicx}
\usepackage{wrapfig}
\usepackage{booktabs}
\usepackage{lineno}
\usepackage{comment}
\usepackage{siunitx}

\newcommand{\Rearth}{$R_\oplus$\xspace}

\newcommand{\gaia}{{Gaia}\xspace}

\newcommand{\kepler}{\emph{Kepler}\xspace}

\newcommand{\ktwo}{\emph{K2}\xspace}

\let\oldmaketitle\maketitle
\renewcommand{\maketitle}{\oldmaketitle\setcounter{footnote}{0}}
\newcommand{\countess}{\texttt{COUNTESS}\xspace}

\newcommand{\UFA}{Department of Astronomy, University of Florida, Gainesville, FL 32611, USA}

\newcommand{\PSUAA}{Department of Astronomy \& Astrophysics, 525 Davey Laboratory, 251 Pollock Road, Penn State, University Park, PA, 16802, USA}
\newcommand{\PSUCEHW}{Center for Exoplanets and Habitable Worlds, 525 Davey Laboratory, 251 Pollock Road, Penn State, University Park, PA, 16802, USA}

\definecolor{plum}{HTML}{88498f}

\begin{document}
\title{COUNTESS I: A Uniformly Vetted Catalog of Known and New Transiting Exoplanets in the TESS Northern Continuous Viewing Zone}
%\title{A Uniformly Vetted and Validated Catalog of Known and Newly Discovered Exoplanets in the TESS North Continuous Viewing Zone}
\shortauthors{Hotnisky et al. 2026}

% \correspondingauthor{Andrew Hotnisky}

\author[0009-0000-1825-4306]{Andrew Hotnisky}
\affiliation{\UFA}
\affiliation{\PSUAA}
\affiliation{\PSUCEHW}
\email[show]{hotniskya@ufl.edu}

\author[0000-0002-3853-7327]{Rachel B. Fernandes}
\altaffiliation{Center for Exoplanets and Habitable Worlds (CEHW) Fellow}
\affiliation{\PSUAA}
\affiliation{\PSUCEHW}
\email{rbf5378@psu.edu}

\author[0000-0003-3702-0382]{Kevin K.\ Hardegree-Ullman}
\affiliation{Caltech/IPAC-NASA Exoplanet Science Institute, 1200 E.\ California Blvd., MC 100-22, Pasadena, CA 91125, USA}
\email{kevinkhu@caltech.edu}

\author[0000-0002-8965-3969]{Steven Giacalone}
\altaffiliation{NSF Astronomy and Astrophysics Postdoctoral Fellow}
\affil{Department of Astronomy, California Institute of Technology, Pasadena, CA 91125, USA}
\email{giacalone@astro.caltech.edu}

\author[0000-0001-8153-639X]{Kiersten M.\ Boley}
\altaffiliation{NASA Sagan Fellow}
\affil{Observatories, Carnegie Institution for Science, Pasadena, CA, 91101, USA}
\email{kboley@carnegiescience.edu}

\author[0000-0001-5847-9147]{Kristo Ment}
\affiliation{\PSUAA}
\affiliation{\PSUCEHW}
\email{kxm821@psu.edu}

\author[0000-0001-9269-8060]{Michelle Kunimoto}
\affil{Department of Physics and Astronomy, University of British Columbia, 6224 Agricultural Road, Vancouver, BC V6T 1Z1, Canada}
\email{mkuni@phas.ubc.ca}

\author[0000-0003-4500-8850]{Galen J.\ Bergsten}
\affil{Space Telescope Science Institute, 3700 San Martin Drive, Baltimore, MD 21218, USA}
\email{gbergsten@stsci.edu}

\author[0000-0002-6673-8206]{Sakhee Bhure}
\affil{Centre for Astrophysics, University of Southern Queensland, Toowoomba, QLD 4350, Australia}
\email{Sakhee.Bhure@unisq.edu.au}

\author[0000-0002-8035-4778]{Jessie L.\ Christiansen}
\affil{Caltech/IPAC-NASA Exoplanet Science Institute, 1200 E.\ California Blvd., MC 100-22, Pasadena, CA 91125, USA}
\email{christia@ipac.caltech.edu}

\author[0000-0002-0015-382X]{Brandon Radzom}
\affil{Department of Astronomy, Indiana University, 727 East 3rd Street, Bloomington, IN 47405-7105, USA}
\affil{Caltech/IPAC-NASA Exoplanet Science Institute, 1200 E.\ California Blvd., MC 100-22, Pasadena, CA 91125, USA}
\email{bradzom@iu.edu}

\author[0000-0001-9596-7983]{Suvrath Mahadevan}
\affiliation{\PSUAA}
\affiliation{\PSUCEHW}
\email{suvrath@astro.psu.edu}

\begin{abstract}
The Transiting Exoplanet Survey Satellite (TESS) has transformed the study of nearby exoplanetary systems; however, its nominal observing strategy limits sensitivity to planets with orbital periods shorter than $\sim$10 days for most parts of the sky. The two TESS Continuous Viewing Zones (CVZs) provide extended temporal baselines that help overcome this limitation, enabling the detection of longer-period ($>$10\,days) transiting planets around nearby stars. Here, we present \countess, a transit-search pipeline optimized for long-baseline TESS observations that combines multi-sector light curves with heterogeneous cadences, and implements fast-folding BLS period detection, vetting, and statistical validation. As a first application of the pipeline, we conducted a search on the primary and first extended mission photometry in the TESS northern CVZ. For this analysis, we used \gaia DR3 and 2MASS photometry to homogeneously derive a stellar catalog of FGKM stars for the TESS northern CVZ, resulting in a sample of 391,059 stars. We used \countess to search for transiting planets around 26,114 of these stars with TESS-SPOC light curves and assessed its performance, recovering 115 out of 159 known TESS Objects of Interest (TOIs; $0.85\ \text{days} < P <124.72\ \text{days}$; $1.03\ R_\oplus < R_p < 16.35\ R_\oplus$). Additionally, we identified 10 new exoplanet candidates ($1.20\ \text{days} < P <34.62\ \text{days}$; $1.73\ R_\oplus < R_p < 4.19\ R_\oplus$) that passed vetting tests, including two new statistically validated sub-Neptunes, TIC~219893931~b and TIC~237254473~b. \countess enables extended-baseline TESS analyses and identification of longer-period planets, establishing a foundation for future exoplanet demographic studies, including comparisons with \kepler and \ktwo.
\end{abstract}

\section{Introduction}
Space-based transit surveys have fundamentally shaped the discovery and characterization of exoplanets over the past two decades. The \kepler mission \citep{borucki_kepler_2010}, which monitored a $\sim$100 square-degree field with a median target distance of $\sim$800\,pc, revolutionized our understanding of planetary systems by delivering the first statistically robust census of thousands of transiting planets spanning a wide range of radii and orbital periods around Sun-like (FGK) stars. Its successor, the \ktwo mission \citep{howell_k2_2014}, extended this legacy with observations of 18 distinct fields along the ecliptic plane for $\sim$80 days each. Together, \kepler and \ktwo revealed consistent population-level structure in the exoplanet distribution, including the radius valley, a deficit of planets near $\sim$1.8\,\Rearth relative to the $\sim$1.3\,\Rearth super-Earth and $\sim$2.4\,\Rearth sub-Neptune populations \citep{fulton_california-kepler_2017,Hardegree-Ullman2020}, and the hot Neptune desert, a region in the period–radius plane ($P < 4\ \text{days}; 3-10$ \Rearth) marked with a scarcity of Neptune-sized planets \citep{beauge_emerging_2013}. However, these surveys primarily targeted distant and relatively faint stellar populations, limiting both their applicability to the solar neighborhood and the feasibility of detailed follow-up mass and atmospheric characterization for many detected planets.

The Transiting Exoplanet Survey Satellite (TESS) \citep{ricker_transiting_2014} was designed to address this limitation by conducting an all-sky survey of bright, nearby stars, thereby enabling the discovery of transiting planets orbiting the closest stars to the Sun. However, TESS’s nominal observing strategy provides only $\sim$27 days of continuous coverage per sector for most of the sky, which biases standard transit searches to planets with orbital periods $\lesssim$10 days when requiring at least two observed transits \citep{kunimoto_predicting_2022, rodel_tiara_2024}.

An important exception is provided by the TESS Continuous Viewing Zones (CVZs), located near the ecliptic poles, where sectors overlap to yield substantially extended observing baselines. \cite{Eschen2024} shows that this extended coverage improves sensitivity to longer-period planets, with non-CVZ TESS observations having lower detection efficiency in period-radius space compared to CVZ observations. In particular, during the Primary Mission (PM; Sectors 1--26), the TESS northern and southern CVZs were observed for 351 days each. During the first Extended Mission (EM1; Sectors 27--55), the TESS southern CVZ was observed for 351 days, and the TESS northern CVZ was observed for 243 days. There is additional coverage of the TESS CVZs during the second Extended Mission (EM2; Sectors 56--83), with generally less full coverage of the northern CVZ (135 days). Furthermore, the third Extended Mission (EM3; Sectors 84--107) is currently being observed with novel pointing strategies such as long stare sectors, rolled sectors, and filling observing gaps toward the galactic center.\footnote{\url{https://heasarc.gsfc.nasa.gov/docs/tess/docs/Guide-to-TESS-for-EM-planning.pdf}} Therefore, we focused on PM and EM1 in this study. Although the continuous coverage in any single TESS observing mission is shorter than \emph{Kepler}’s four-year baseline, the total multi-cycle temporal coverage rivals that of \kepler and is greater than \ktwo ($\sim$80 days per campaign), enabling sensitivity to significantly longer-period planets and improved transit parameter precision through the accumulation of additional transit events. As such, the TESS CVZs provide a unique bridge between the distant, well-characterized \kepler population 
%($d_{\rm median}\approx800$ pc) 
and nearby planetary systems accessible to detailed follow-up. 
%($d_{\rm median}\approx219$ pc).

However, the CVZs introduce several challenges not encountered in standard short-baseline TESS analyses. Most notably, the photometric cadence changes from 30 minutes in the PM Full Frame Images (FFIs) to 10 minutes in EM1, altering both transit morphology and noise properties across the observing baseline. These heterogeneous cadences complicate the coherent detection of transit signals across multiple sectors and mission phases. While several TESS transit-search pipelines exist \citep[e.g.,][]{feliz_nemesis_2021,boley_searching_2021,fernandes_pterodactyls_2022,gan_occurrence_2022}, no current framework is specifically designed to address all of these CVZ and multi-mission-specific challenges in a homogeneous, end-to-end manner. A dedicated and uniform CVZ pipeline is essential not only for detecting long-period planets but also for enabling the characterization of completeness and the measurement of occurrence rates required for robust exoplanet demographics. Such a framework would enable direct comparisons between the nearby TESS planet population and the demographic trends identified by \kepler and \ktwo, while also yielding bright, nearby long-period planet candidates well suited for mass and atmospheric follow-up.

Here, we present \countess (\textbf{C}ombining \textbf{O}bservations to \textbf{U}nveil \textbf{N}ew \textbf{T}ransiting \textbf{E}xoplanet \textbf{S}ystems and \textbf{S}tatistics), a pipeline that builds on publicly available and tested tools to extract, detrend, search, vet, and statistically validate transiting planet candidates detected in the TESS CVZs using combined 30-minute PM and 10-minute EM1 cadence FFIs. In this paper, as a proof of concept, we focused on the TESS northern CVZ. In Section~\ref{sec:stellar}, we discuss how we used data from \gaia DR3 to select our target sample and additional data from 2MASS to homogeneously derive stellar effective temperatures, luminosities, radii, and masses.  Section~\ref{sec:Pipeline} details the data processing, transit search, model-fitting, vetting, and validation procedures implemented in \countess. In Section~\ref{sec:Results}, we discuss the recovery of known TOIs and the discovery of new planet candidates. In Section~\ref{sec:Summary}, we summarize our results, the new planet candidate discoveries, and explore future work and demographic studies with \countess.

\section{Stellar Sample} \label{sec:stellar}

Because the measured exoplanet transit depth ($\delta$) is proportional to the square of the ratio of the radius of the planet ($R_p$) to the radius of the star ($R_\star$), $\delta \propto (R_p/R_\star)^2$, the precision with which we can measure an exoplanet's radius explicitly depends on the stellar radius measurement. For exoplanet demographic studies, it is critical to have a uniformly derived stellar parameter set in order to identify trends not only with respect to planet radii (and masses if working with radial velocity data), but also with respect to host star properties (e.g., spectral type/effective temperature, metallicity). We adopted the same stellar characterization methodology of \citet{Hardegree-Ullman2025} in order to make the TESS northern CVZ dataset compatible with the \kepler and \ktwo datasets for future combined exoplanet demographics studies. This stellar characterization methodology has been fine-tuned for exoplanet demographics studies and builds upon previous work outlined in \citet{Hardegree-Ullman2019,hardegree-ullman_scaling_2020,hardegree-ullman_bioverse_2023}, and \citealt{Fernandes2023}. We briefly describe the methodology used to derive stellar parameters here.

\subsection{Initial Target Sample}

We downloaded and concatenated the TESS Input Catalog files\footnote{\url{https://archive.stsci.edu/tess/tic_ctl.html}} \citep[v8.2;][]{Stassun2019,Paegert2021} for declinations above 54 degrees, which contained 79,951,213 targets. To narrow the targets to the TESS northern CVZ, we selected targets via a cone search within a 12$^{\circ}$ radius of the northern ecliptic pole centered at 18\si{h}00\si{m}00\si{s}, \ang{+66;33;38.55}, which resulted in 3,907,252 targets. To obtain the latest \gaia photometry, we identified \gaia DR3 IDs for each TIC target by performing a cross-match of our TIC targets with \gaia DR3\footnote{\url{https://github.com/kevinkhu/ticgaia}}. For stellar characterization, we require distances and full photometry, so we dropped targets without distances from \citet{Bailer-Jones2021} and full $G$, $G_{BP}$, and $G_{RP}$ and 2MASS \citep{Skrutskie2006} $J$, $H$, and $K$-band photometry, leaving us with 1,269,074 targets. The majority of targets removed in this step did not have 2MASS photometry, and were generally fainter than a $G_{RP}$ magnitude of 17, for which we do not expect to yield any detectable exoplanet transit signals. The $G_{RP}$ bandpass is similar to the TESS bandpass between $\sim$600 and 1000\,nm \citep{ricker_transiting_2014,Montegriffo2023}, so we used it as a proxy for the TESS-band magnitude. Additionally, the photometric precision of TESS diminishes as a function of magnitude, so we limited our sample to stars with $G_{RP}<16$, leaving 591,444 stars.

Our stellar parameter derivations below assume our targets are single stars. While we cannot remove all stellar multiples from a sample this large, we can mitigate some contamination by limiting our targets to a \gaia RUWE score $<1.4$ \citep{Lindegren2018} and a \gaia \texttt{non\_single\_star} flag equal to 0 \citep{GaiaCollaboration2023}, which left us with 530,514 targets. Lastly, we made a quality cut for distances $<5000$~pc and distance uncertainties $<10\%$, which brought the total initial stellar sample to 496,585 targets. These distance quality cuts were imposed for uniformity with the TESS southern CVZ sample (S. Bhure et al. in prep) and they largely mitigate contamination from the Large Magellanic Cloud in the southern CVZ. We note that only 0.2\% of TESS northern CVZ stars are beyond 3000~pc, and these distant targets are fainter than $G_\mathrm{RP}=14$, so they are unlikely to observationally detect exoplanets.

\subsection{Stellar Parameters}

Here we describe how we computed key stellar properties for our sample including, stellar effective temperature $T_\mathrm{eff}$, surface gravity $\log g$, metallicity [Fe/H], luminosity $L_\star$, radius $R_\star$, and mass $M_\star$. A color-temperature relationship was adopted to compute $T_\mathrm{eff}$ from a 4th-order polynomial fit to $T_\mathrm{eff}$ versus \gaia $G_{BP}-G_{RP}$ color from the individual stellar data used to compile Table 5 of \citet{Pecaut2013}.\footnote{We used updated stellar data maintained at \url{https://github.com/emamajek/SpectralType}, accessed June 2025.} 

Metallicity and a preliminary surface gravity were computed using random forest regression \citep{Pedregosa2011} trained on \gaia and 2MASS colors and extinction-corrected \citep[3D Bayestar19 dust map, \texttt{dustmaps} code][]{Green2018,Green2019}, and 2MASS absolute magnitudes from a subset of 9,043 TESS northern CVZ targets that had [Fe/H] and $\log g$ measurements from LAMOST Data Release 11 \citep[\url{https://www.lamost.org/dr11/};][]{Cui2012}.

Bolometric luminosity was computed from the $K$-band bolometric magnitude, applying a bolometric correction computed using \texttt{isoclassify} \citep{Huber2017}, which interpolates $T_\mathrm{eff}$, $\log g$, and [Fe/H] onto the MIST stellar model grid \citep{Choi2016}. Effective temperature and the bolometric luminosity were input into the Stefan-Boltzmann law to compute radii for FGK stars, and the empirical radius-luminosity relationship of \citet{Mann2015} was used for M stars in the range $4.5 < M_{K_s} < 10$.

For the same magnitude range, masses for M dwarfs were computed using the mass-luminosity relationship of \citet{Mann2019}. Masses for FGK stars were computed using the mass-luminosity relationship from \citet{Torres2010}. An updated surface gravity was then computed directly from the mass and radius values.

\begin{figure}[ht]
    \centering
    \includegraphics[width=\columnwidth]{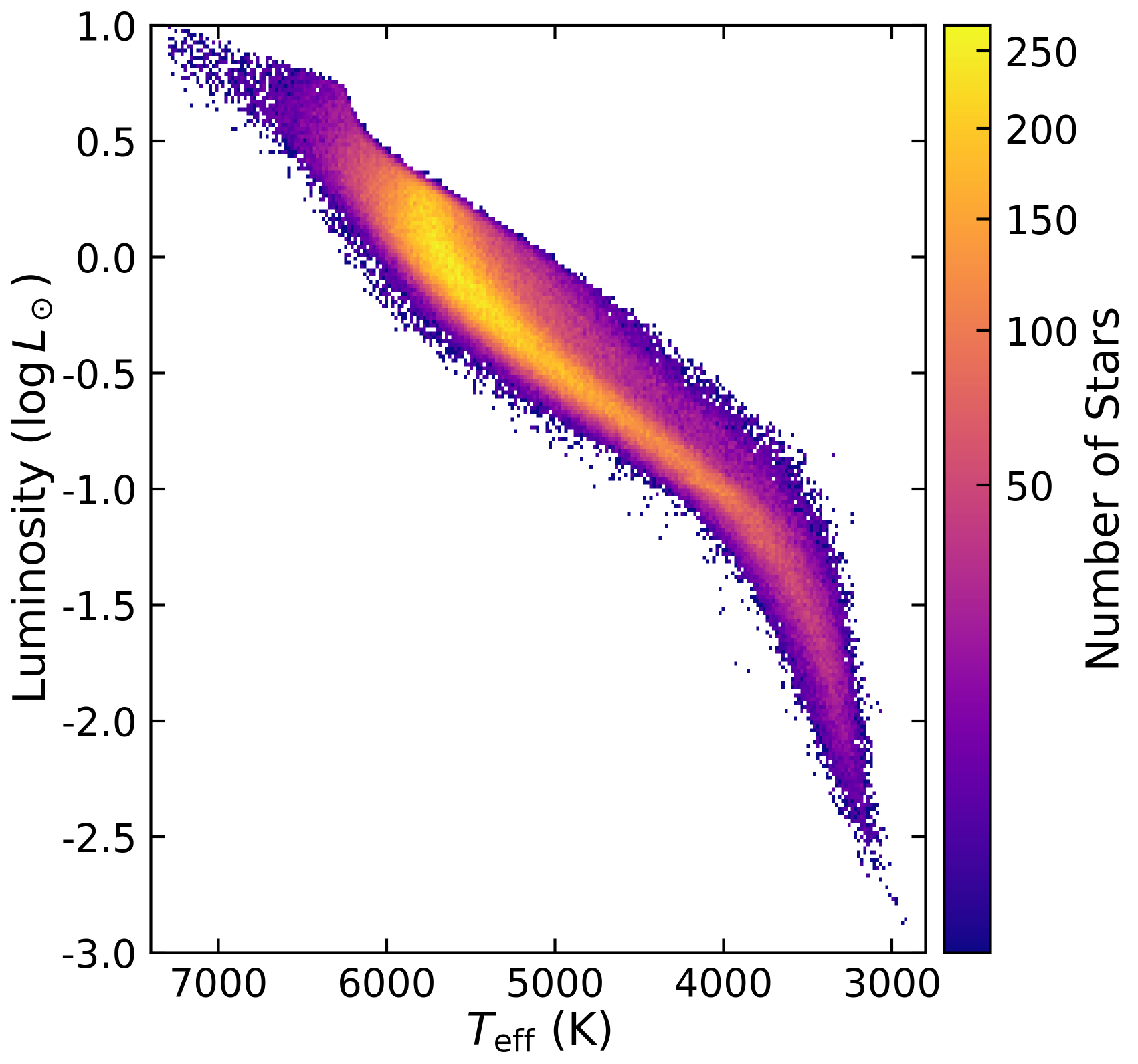}
    \caption{H-R diagram of the TESS northern CVZ stellar sample. The colors on the diagram indicate the number density of stars in our sample.}
    \label{fig:HR}
\end{figure}

Lastly, we limited our sample to main sequence FGKM stars with surface gravity in the range of non-evolved stars \citep[see Equation 9 of][]{Huber2016}, which left 383,718 stars. Figure~\ref{fig:HR} shows a Hertzsprung-Russell (H-R) diagram of the TESS northern CVZ stellar sample. All stellar properties are given in Table~\ref{tab:stars}.

\begin{deluxetable*}{clccrccc}[htb!]
\tablecaption{Uniformly Derived Properties of TESS northern CVZ Stars\label{tab:stars}}

\tablehead{\colhead{TIC ID} & \colhead{Distance} & \colhead{$T_{\mathrm{eff}}$} & \colhead{$\log g$} & \colhead{[Fe/H]} & \colhead{$L_{\star}$} & \colhead{$R_{\star}$} & \colhead{$M_{\star}$} \vspace{-0.75em}\\ 
\colhead{} & \colhead{(pc)} & \colhead{(K)} & \colhead{[cm s$^{-2}$]} & \colhead{(dex)} & \colhead{($L_{\odot}$)} & \colhead{($R_{\odot}$)} & \colhead{($M_{\odot}$)}
} 

\startdata
160491047 & $877.75_{-22.84}^{+22.47}$ & $4672\pm137$ & $4.559\pm0.071$ & $-0.203\pm0.233$ & $0.215\pm0.021$ & $0.711\pm0.053$ & $0.680\pm0.055$ \\
160491087 & $934.11_{-10.33}^{+11.91}$ & $5987\pm175$ & $4.312\pm0.070$ & $-0.439\pm0.233$ & $1.711\pm0.141$ & $1.220\pm0.085$ & $1.114\pm0.091$ \\
160491110 & $1007.99_{-25.52}^{+24.15}$ & $4590\pm135$ & $4.397\pm0.062$ & $-0.088\pm0.233$ & $0.325\pm0.028$ & $0.908\pm0.056$ & $0.749\pm0.061$ \\
160493787 & $977.98_{-33.08}^{+29.47}$ & $4760\pm140$ & $4.486\pm0.100$ & $0.100\pm0.233$ & $0.279\pm0.033$ & $0.792\pm0.071$ & $0.723\pm0.061$ \\
160493795 & $187.20\pm0.40$ & $6209\pm182$ & $4.149\pm0.074$ & $-0.158\pm0.233$ & $3.314\pm0.247$ & $1.572\pm0.109$ & $1.309\pm0.105$ \\
160493807 & $346.11_{-1.46}^{+1.09}$ & $5195\pm152$ & $4.454\pm0.067$ & $0.250\pm0.233$ & $0.562\pm0.041$ & $0.916\pm0.064$ & $0.857\pm0.069$ \\
160493808 & $1187.14_{-57.34}^{+53.47}$ & $4898\pm144$ & $4.553\pm0.078$ & $-0.032\pm0.233$ & $0.304\pm0.033$ & $0.752\pm0.068$ & $0.737\pm0.061$ \\
160493810 & $1439.20_{-68.84}^{+63.15}$ & $5319\pm156$ & $4.572\pm0.099$ & $0.033\pm0.233$ & $0.449\pm0.074$ & $0.784\pm0.076$ & $0.806\pm0.071$ \\
160493819 & $1089.63_{-36.32}^{+45.40}$ & $5152\pm151$ & $4.608\pm0.071$ & $-0.187\pm0.233$ & $0.296\pm0.039$ & $0.698\pm0.055$ & $0.732\pm0.062$ \\
160493821 & $1333.28_{-55.12}^{+61.53}$ & $5516\pm162$ & $4.505\pm0.077$ & $-0.199\pm0.233$ & $0.593\pm0.072$ & $0.854\pm0.067$ & $0.863\pm0.072$ \\
\enddata

\tablecomments{Only a portion of this table is shown here to demonstrate its form and content. A machine-readable version of the full table is available.}
\end{deluxetable*}

%To further refine our stellar sample for comparisons to demographic studies of FGKM dwarfs, we only included stars with Teff in the range of 2300–7300 K and log g in the range of nonevolved stars based on the dwarf criterion from D. Huber et al. (2016; their Equation (9)), leaving 108,019 Kepler stars and 133,848 unique \ktwo stars. We also excluded probable binary stars with Gaia RUWE values greater than 1.4 and Gaia non_single_star not equal to 0, leaving 95,092 Kepler stars and 113,402 unique \ktwo stars. Targets with noisy photometry provide little constraining power and merely slow down the calculations, therefore, we removed targets with CDPP8hr > 1200 ppm for \ktwo stars, as was done in previous Scaling \ktwo papers (J. K. Zink et al. 2020a, 2020b, 2023; J. L. Christiansen et al. 2023), and CDPP7.5hr > 1000 ppm for Kepler stars. These cuts left 94,422 Kepler and 90,344 unique \ktwo FGKM main-sequence stars. Properties for the entire Kepler and \ktwo stellar samples are giv

\section{\countess} \label{sec:Pipeline}
\begin{figure*}
    \centering
    \includegraphics[width=1.00\textwidth]{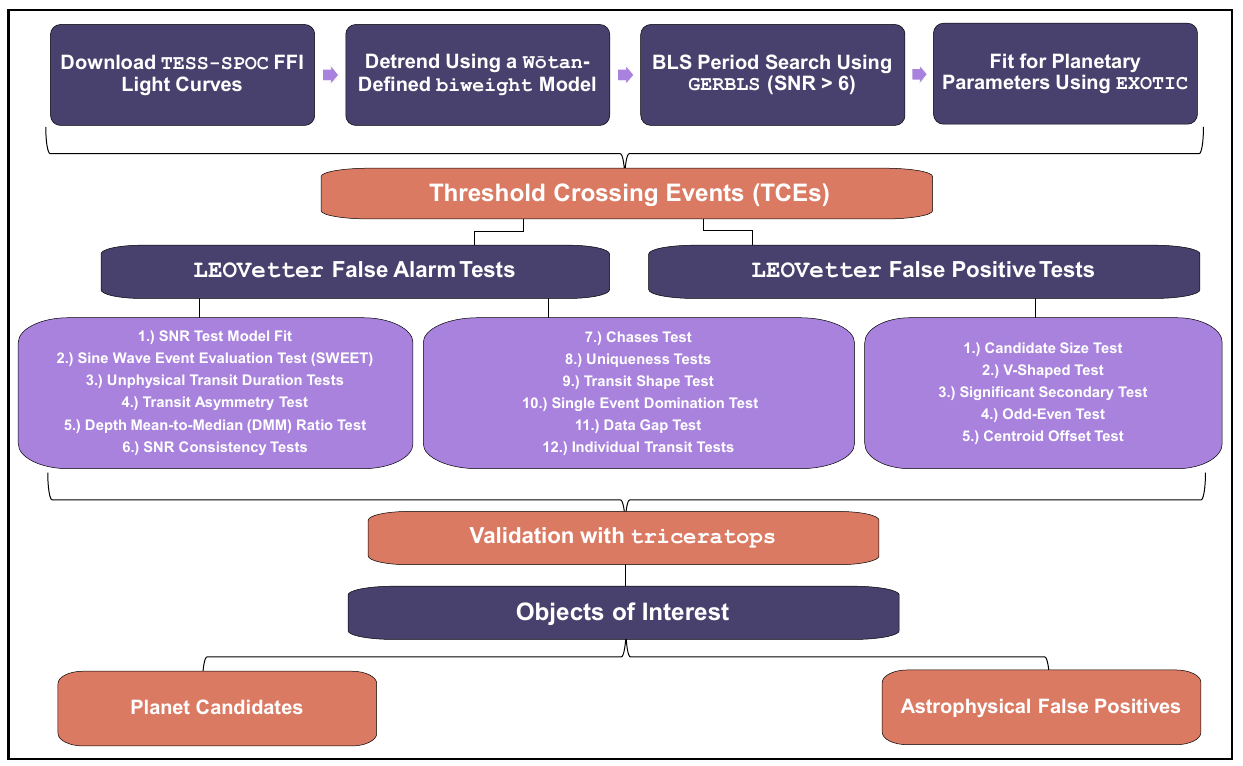}
    \caption{The schematic flow chart of how a target is processed through \countess.}
    \label{fig:Flow}
\end{figure*}

We focused on the TESS northern CVZ, combining PM and EM1 FFI data, which provide longer baselines but heterogeneous cadences (PM: 30-minute; EM1: 10-minute) and differing noise properties. To address these challenges, we developed \countess a unified end-to-end workflow optimized for long-baseline, multi-cadence TESS CVZ light curves, enabling the uniform searching, vetting, and statistical validation of long-period planet candidates. This process is outlined in Figure~\ref{fig:Flow}. In the following sections, we describe each stage of \countess in detail.

\subsection{Light Curve Download} \label{sec:lcdownload}
Unlike the 2-minute pre-selected sample of TESS targets, we used the FFI data, which provide an unbiased, larger stellar sample for planet searches and demographic studies. We began by downloading all available TESS Science Processing Operations Center PDCSAP \citep[SPOC;][]{jenkins_tess_2016} FFI light curves using the \texttt{lightkurve} package \citep{lightkurve_collaboration_lightkurve_2018}. The TESS-SPOC products are limited by selection criteria that prioritize nearby, bright stars ($T_{\text{mag}} \leq 13.5$). Here, we restricted our analysis to data from PM and EM1 to establish our long-period detection framework, which we will expand to EM2 in future work. We only preserved sectors with FFI \texttt{lightkurve} bitmask quality value equal to 0. This selection yielded long-baseline light curves, with PM and EM1 coverage suitable for our blind search, of 26,114 TESS northern CVZ stars. The targets in our sample span a range of downloadable sector coverage, from 1 to 24 sectors, with a median of 22 sectors per target. While we initially explored the use of light curves from the Quick Look Pipeline \citep[QLP;][]{huang_photometry_2020} and the TESS-\gaia Light Curves \citep[TGLC;][]{han_tess-gaia_2023}, we found that the TESS-SPOC light curves (median RMS $\approx 2500$ ppm) had lower noise properties compared to QLP (median RMS $\approx 5500$ ppm) and TGLC (median RMS $\approx 12200$ ppm), thereby allowing for better planet recovery.

The cadence of the FFIs changed from 30 minutes in the PM to 10 minutes in EM1. Upon testing, we found that preserving the native cadences of each mission leads to heterogeneous sampling and noise properties across the combined baseline, thereby reducing the precision of recovered orbital periods and planet radii. To mitigate this issue, we binned the 10-minute EM1 light curves to a 30-minute cadence to match the PM data. We also tested \countess on available 2-minute SPOC light curves; however, the 2-minute data did not recover missed signals, as the increased number of data points was insufficient to overcome the higher per-point scatter in these light curves. The 2-minute light curves produced performance broadly consistent with our current binned-cadence pipeline when using the same light-curve products. Binning reduces the point-to-point scatter in EM1 data and ensures consistent noise properties within a fixed time-based detrending window, allowing for consistent baseline structure. Although binning can remove very short-duration transit structure, our search is focused on multi-sector, long-baseline signals, where matching the cadence and stabilizing the noise behavior across mission phases leads to more reliable transit recovery. We then concatenated the light curves across sectors and missions to produce a single, continuous time baseline for detrending and transit searches (Sections~\ref{sec:lcdetrend} and~\ref{sec:GERBLS}). 

%Each light curve is downloaded into the following format: sector number, time, flux, flux error, and cadence value. This allows us to 
%Most targets retain coverage across the majority of PM and EM1 sectors, with a small number of sectors excluded due to instrumental artifacts. We only preserve sectors with FFI lightkurve bitmask quality value equal to 0.

\subsubsection{Accounting for systematics} \label{sec:system}
To ensure that the transit signals identified in our search are astrophysical and not driven by data/instrumental artifacts, we applied a  diagnostic similar to the “Skye” excess metric from \kepler\ detailed in \cite{thompson_planetary_2018}, which has now been used for \ktwo \citep[e.g.,][]{zink_scaling_2020} and TESS \citep[e.g.,][]{fernandes_pterodactyls_2022}. This test measures the number of transit signals at each time stamp and flags instances where the number of detections exceeds the 3$\sigma$ threshold of period signals in any given sector. We ran our 26,114 TESS-SPOC target light curves through \countess and applied this metric. In sectors 14, 16, 17, 18, 19, and 22, we identified time stamps with anomalously high numbers of detected transit events (see Figure~\ref{fig:SME} in Appendix~\ref{app:sem}). We then masked these time stamps and reran the transit search.

\begin{figure*}
    \centering
    \includegraphics[width=\textwidth]{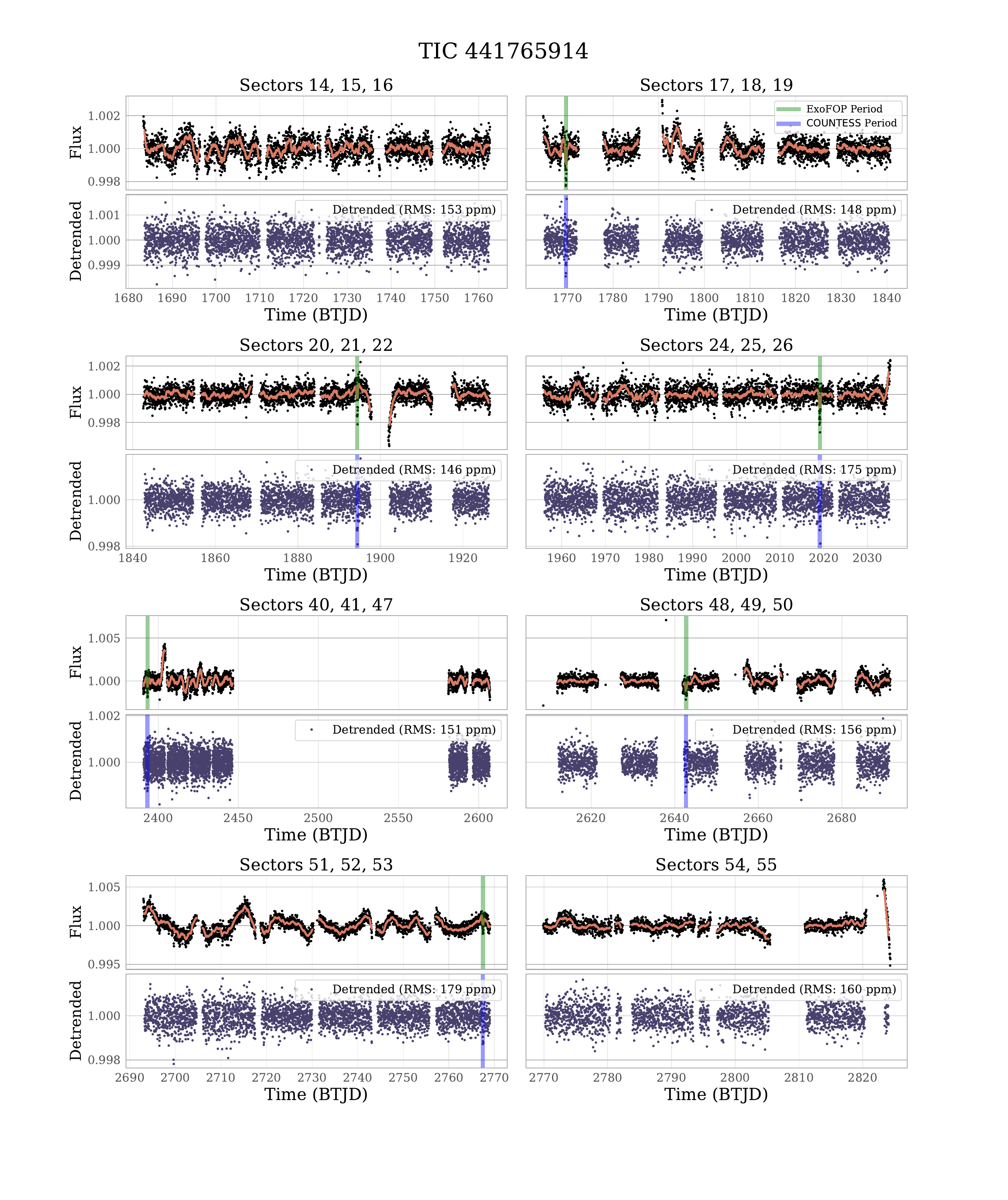}
    \vspace{-50pt}
    \caption{A sector-by-sector detrending of TIC 441765914 (TOI-2088). The top panel of each sector shows the raw flux (black) from the TESS-SPOC FFIs with a detrending curve (orange). The bottom panel shows the detrended light curve. The transits computed from the ExoFOP epoch and from the period recovered by \countess are displayed as vertical bars in green (ExoFOP) and blue (\countess).}
    \label{fig:TOI2088}
\end{figure*}

\begin{figure}[ht]
    \centering
    \includegraphics[width=\columnwidth]{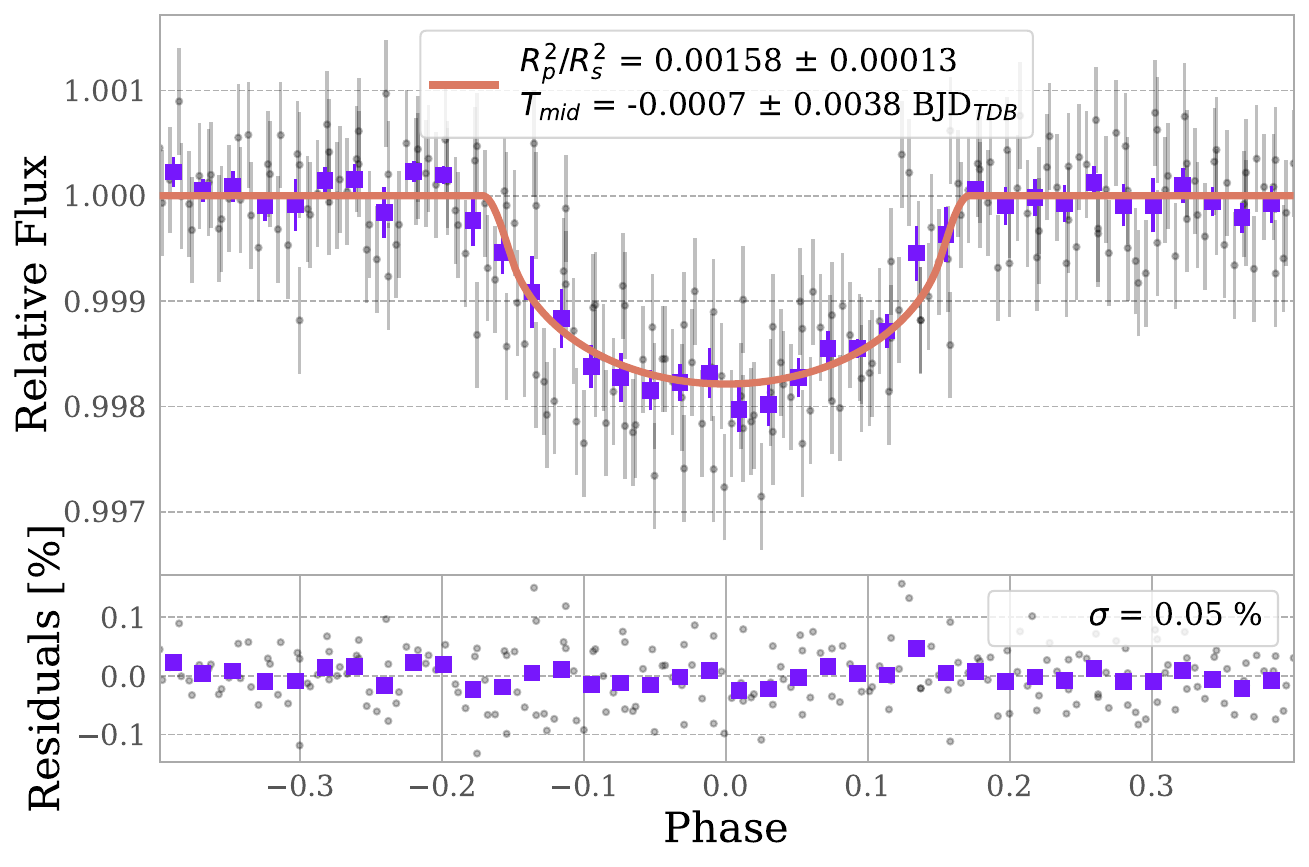}
    \caption{The upper panel shows a phase-folded light curve of TOI-2088 b, with the best-fit model in red from \texttt{EXOTIC}. The lower panel shows the residuals in \%.}
    \label{fig:TOI2088_exotic}
\end{figure}

\subsection{Light Curve Detrending} \label{sec:lcdetrend}
Stellar variability in light curves can obscure genuine transit signals or introduce spurious, transit-like features that hinder our search for planets \citep{krenn_detecting_2024}. To mitigate stellar variability, we tested multiple detrending approaches implemented in the \texttt{W\={o}tan} package\footnote{\url{https://github.com/hippke/wotan}} \citep{hippke_wotan_2019}, including spline-based methods and Gaussian Process (GP) regression. We evaluated these detrending methods using the known TOI sample by comparing whether the known TOI periods were recovered and whether the recovered transit depths were preserved after detrending. We compared the recovered periods and radii obtained with the biweight, spline, and GP detrending models and used the method that yielded the highest total number of successfully recovered TOIs. Following this criterion, we found that the biweight filter \citep{Mosteller1977} most reliably removed the stellar variability while preserving transit depths across our sample. Additionally, \cite{hippke_wotan_2019} found that the biweight filter was shown to be the best-performing detrending model for \kepler data.

% Therefore, we used the biweight filter as implemented in the\texttt{W\={o}tan} package\footnote{\url{https://github.com/hippke/wotan}} \cite{hippke_wotan_2019} to detrend the light curves. This time-windowed filter is designed to preserve transit signals while removing stellar variability. Motivated by the results found in \cite{hippke_wotan_2019}, we tested the biweight filter alongside a Gaussian Process (GP) detrending approach and found that the biweight filter more reliably preserved transit depths and shapes across the TESS northern CVZ TOI sample. 

A critical parameter in the biweight filter is the window length, defined as the duration of the sliding time window used to estimate and remove the underlying flux trend, thereby setting the characteristic timescale of variability that is filtered out. If the window length is too short, long-duration transit signals will be partially or fully removed. If the window length is too long, the stellar variability will not be subtracted properly, preserving possible false periodic signals. We first tested a window length of 0.5 days \citep[as recommended by][]{hippke_wotan_2019} on the light curves of known northern CVZ TOIs. We found that while this window length reliably recovered short-period transit signals, a short window length of 0.5 days led to over-detrending, which partially removed transit signals (see Figure~\ref{fig:DetrendComp} in Appendix~\ref{app:WLF}). Because the orbital periods are not known a priori, the choice of window length cannot be tailored to individual systems at this stage. To address this, we implemented a two-stage window-length approach. All light curves were first detrended using a baseline window length of 0.5 days, and the transit search was then performed on the detrended data to recover candidate orbital periods. For candidates with recovered orbital periods exceeding 100 days, we discarded the initial detrending and, instead, reprocessed the light curves using an extended window length of 1.75 days, which we found best mitigated over-detrending of the transit signal. Essentially, our window length $w$ as a function of period $P$ is:

% we re-ran the search using a longer detrending window to ensure that the transit depth and duration were not artificially decreased by over-aggressive detrending. The window length ($w$) was therefore defined as:

\begin{equation*}
    w(P) = \begin{cases}
        0.5 \text{ days} & \text{if } P \leq 100~\text{days} \\
        1.75 \text{ days} & \text{if } P > 100~\text{days}.
    \end{cases}
\end{equation*}
The longer window length reduces the over-detrending by extending the baseline over which the local trend is estimated, thereby preserving long-duration transit structure. This adjustment improves the accuracy of the inferred transit depth and duration, and therefore the derived planetary radius for long-period signals while maintaining recovery of shorter-period signals.  

\subsection{\texttt{GERBLS} Period Search} \label{sec:GERBLS}
We searched the detrended light curves for periodic, transit-like signals using the Greatly Expedited Robust Box Least Squares (\texttt{GERBLS}; Ment et al. submitted)\footnote{\url{https://github.com/kment/GERBLS}} code. Like a traditional Box-Least Squares (BLS), \texttt{GERBLS} models transits as simple box-shaped dips in flux. Instead of brute-force phase-folding the light curve over tens to hundreds of thousands of trial periods and phases, it employs a fast-folding algorithm and a divide-and-conquer strategy that reuses intermediate computations. This approach reduces the runtime of an individual BLS search by roughly an order of magnitude while enabling a comparable increase in the number of trial periods tested, substantially accelerating large transit searches and improving period resolution without the need for a GPU.

For \countess, we adopted a signal-to-noise ratio (S/N) threshold of 6 for all Threshold Crossing Events (TCEs), computed using the \texttt{GERBLS} S/N definition, consistent with other TESS surveys \citep[e.g.,][]{feliz_nemesis_2021, fernandes_pterodactyls_2022, hadjigeorghiou_raven_2025}. \texttt{GERBLS} recovers the transit epoch, period, and approximate depth that are passed to the transit fitting. 

\subsection{Transit Fitting} \label{sec:TransitFit}

\texttt{GERBLS} only provides a first-order approximation of the transit parameters, i.e., it does not capture ingress and egress structure or limb-darkening effects, and therefore cannot fully constrain the transit geometry. As a result, an additional transit-fitting step is required to derive physically meaningful parameters for planet candidates. We fit for the transit parameters using the EXOplanet Transit Interpretation Code \citep[\texttt{EXOTIC};][]{zellem_utilizing_2020}\footnote{\url{https://github.com/rzellem/EXOTIC}}. This open-source Python package employs UltraNest Bayesian nested sampling \citep{buchner_ultranest_2021} to converge on the most probable transit model roughly twice as fast as a typical MCMC. We also tested the \texttt{batman} \citep{kreidberg_batman_2015} and \texttt{exoplanet} \citep{foreman-mackey_exoplanet_2021, foreman-mackey_exoplanet-devexoplanet_2021} packages alongside \texttt{EXOTIC}. We adopt the \texttt{EXOTIC} fitting routine, as it offers the lowest computational cost while returning parameter estimates consistent with those obtained from the other fitting methods.

We passed the \texttt{GERBLS} phase-folded light curve to \texttt{EXOTIC} as the data to be fit. The \texttt{GERBLS}-derived period, transit duration, and $R_p/R_\star$ estimates were used as priors for the \texttt{EXOTIC} fit. We fit for $R_p/R_\star$, $a/R_\star$, $i$, and limb darkening coefficients $u_0$ and $u_1$. The bounds on these parameters encompass the physically plausible range of planet sizes and orbital configurations per star in our sample while excluding clearly nonphysical solutions. To further validate these thresholds, we visually inspected the fitted transit models and residuals to ensure that the models traced the data and reproduced the expected transit geometry in the detrended light curves.

\subsection{Candidate Vetting \& Statistical Validation} \label{sec:LEOVet}

Various artifacts can mimic planetary transit-like signals in TESS light curves, which fall into two categories: false alarms (FAs) and astrophysical false positives (FPs). FAs are signals that are introduced by instrumental noise and data artifacts such as scattered light features and momentum dumps. FPs are any transit events that are not caused by a planet, but rather other astrophysical phenomena such as eclipsing binaries (EBs) and background EBs. Therefore, we need to vet and statistically validate any transit signals we find to distinguish the planetary transits from FAs and FPs.

We vetted all the TCEs using \texttt{LEOVetter}\footnote{\url{https://github.com/mkunimoto/LEO-vetter}} \citep{kunimoto_leo-vetter_2025}, which builds upon the \texttt{RoboVetter} framework developed for the \kepler mission \citep{thompson_planetary_2018}. \texttt{LEOVetter} evaluates each TCE using a list of diagnostic tests designed to identify both instrumental and astrophysical false signals. In total, \texttt{LEOVetter} applies 12 FA tests and 5 FP tests. The FA tests assess signal consistency, including checks on S/N and transit depth across individual events, while the FP tests include examining odd–even transit depth differences, transit alignment, and whether the inferred companion size is consistent with a planetary object. The complete list of tests is shown in Figure~\ref{fig:Flow}, with the corresponding thresholds listed in Table~\ref{tab:LEOThresh} in Appendix~\ref{app:APPVet}. Additionally, \texttt{LEOVetter} includes a pixel-level vetting step in which it uses TESS's Target Pixel File (TPF) to evaluate whether the source of the transit signal is spatially consistent with the target star, allowing us to identify cases where the signal originates from a nearby contaminating source rather than the intended target. 

The TCEs that pass vetting are then processed using \texttt{triceratops}\footnote{\url{https://github.com/stevengiacalone/triceratops}} \citep{giacalone_vetting_2021}, a statistical validation tool constructed to distinguish astrophysical false positives from reliable planet candidates in TESS data. \texttt{triceratops} computes the marginal likelihood of a set of competing transit-producing scenarios using parameters derived from the transit model and the local stellar environment, such as the measured transit depth and orbital period and the properties and positions of nearby stars from \gaia DR3 \citep{GaiaCollaboration2023}. The scenarios evaluated by \texttt{triceratops} include, but are not limited to, EBs associated with the target star, EBs with a known nearby star, EBs with an unresolved background star, and a transiting planet associated with the target star. Using these likelihoods, \texttt{triceratops} calculates a false-positive probability (FPP) and nearby false-positive probability (NFPP). \cite{giacalone_vetting_2021} lists that ``likely planets'' have an FPP $<$ $0.5$ and NFPP $<$ $10^{-3}$. We adopted these values to assess whether our results are planet candidates or false positives. To confirm that our resulting FPP and NFPP values were accurate, we ran each target through the validation 20 times and took the median of the FPPs and NFPPs.

%Maybe Add in a table of all the thresholds of each test? Or do I explain each test in detail? Can do either relatively easily

\section{Results} \label{sec:Results}

\begin{figure*}[t]
    \centering
    \begin{minipage}[t]{0.95\textwidth}
        \centering
        \includegraphics[width=\textwidth]{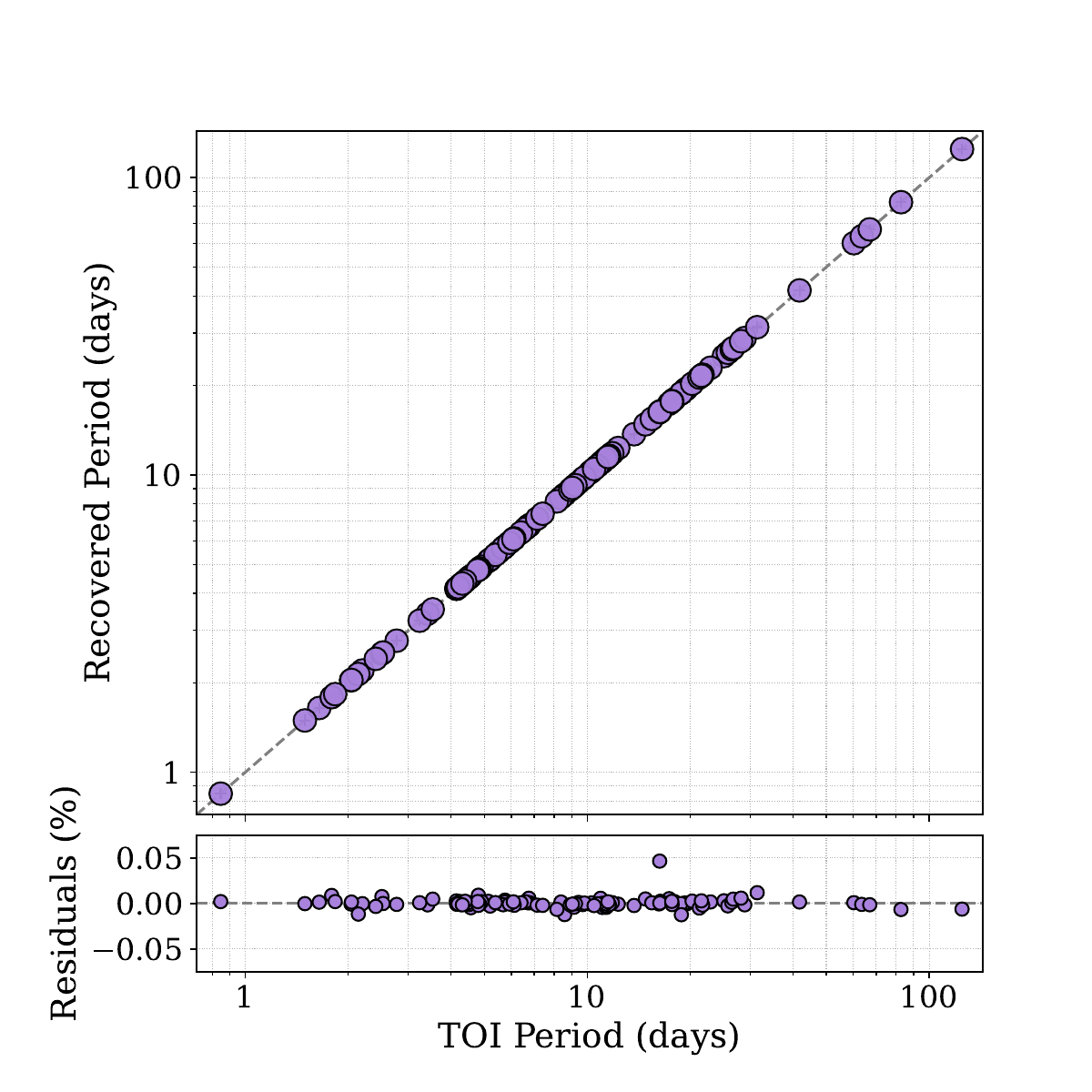}
    \end{minipage}
    \hspace{0.005\textwidth}
    \caption{The comparison plots of recovered TOIs for the ExoFOP measured orbital periods compared to the \countess discovered periods.}
    \label{fig:1to1P}
\end{figure*}

\begin{figure*}[t]
    \centering
    \begin{minipage}[t]{0.95\textwidth}
        \centering
        \includegraphics[width=\textwidth]{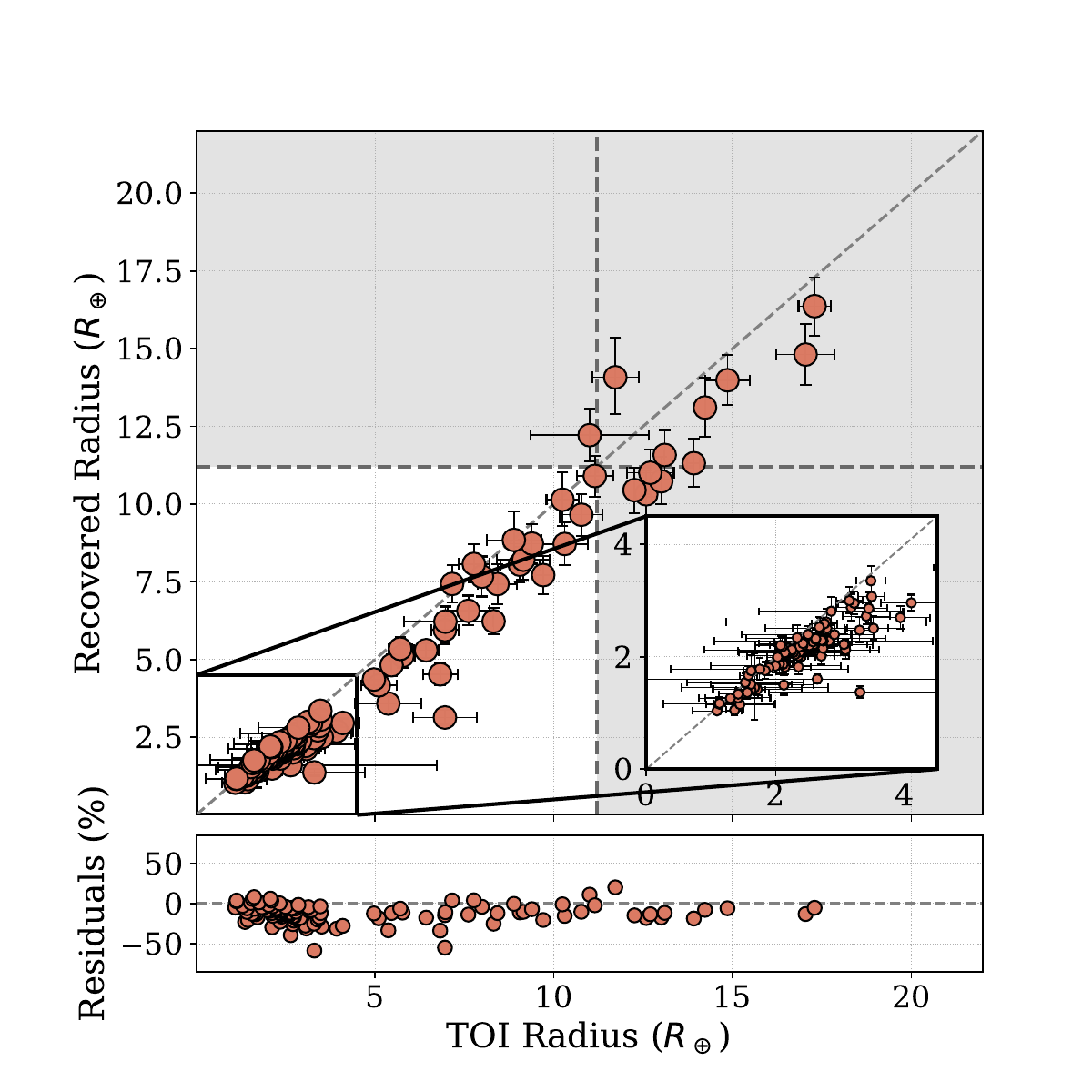}
    \end{minipage}
    \caption{The comparison plot of recovered TOIs for the ExoFOP measured planet radii compared to the \countess discovered planet radii. Note that the planetary parameters from ExoFOP are not homogeneously determined, resulting in expected scatter in the radius fit.}
    \label{fig:1to1}
\end{figure*}

We searched for transiting planets in the PM and EM1 light curves of 26,114 FGKM stars in the TESS northern CVZ using \countess, our automated transit-search and vetting pipeline optimized for multi-sector TESS full-frame image photometry. As an illustrative example, we show how \countess handles the detrending and recovery of TOI-2088 b in Figures~\ref{fig:TOI2088} and~\ref{fig:TOI2088_exotic}. 
%Figure~\ref{fig:1to1} shows the period and planet radius recovery of \countess in comparison to catalog values on ExoFOP. Figure~\ref{fig:RadvPer} details the period and planetary radii distribution of recovered TOIs and new candidates.

\subsection{Recovery of Known TOIs} \label{sec:recov}

We used a sample of 191 TESS northern CVZ TOIs observed during the PM and EM1, drawn from ExoFOP\footnote{Accessed May 3$^{\text{rd}}$ 2026. \url{https://exofop.ipac.caltech.edu/tess/view_toi.php}} \citep{christiansen_nasa_2025}, to train and test the planet recovery, vetting, and statistical validation modules of \countess. Of these 191 TOIs, 115 have the TFOP disposition of planet candidates (PC), 44 confirmed or known planets (CP/KP), 9 FAs, and 23 FPs. We list the TOIs in Table~\ref{tab:TOIs} in Appendix~\ref{app:TOIs}. Following the methodology described in Section~\ref{sec:Pipeline}, we searched for each TOI within its combined PM$+$EM1 light curves, adjusting the transit-fitting and vetting thresholds of \countess to recover the known PCs, KPs, and CPs while removing the FAs and FPs. The TOIs span orbital periods ranging from 0.27 days to 833 days. To cover this full orbital period range, we ran the TOI targets through \countess twice: first with a window length of 0.5\,days for all targets, and then with a window length of 1.75\,days for $P > 100$. This allowed for not only the recovery of the PCs, CPs, and KPs, such as TOI-2088~b, but also identifying FPs with large radii which had previously seemed like PCs due to the detrending removing part of the transit depth.

\countess successfully flagged all TOIs that were known to be FAs and FPs. However, we only recovered 115 of the 159 PCs, CPs, and KPs. Of the 44 TOIs missed by \countess, 38 are PCs, and 8 are CPs/KPs. Most were missed at the BLS stage: 29/44 did not pass our BLS recovery criteria, either because the recovered period did not match the TOI period or because the signal fell below our adopted SNR threshold (S/N $<$ 6). The remaining 15/44 were detected by BLS but were rejected during vetting, commonly due to inconsistent transit SNRs, non-unique events, significant secondaries, off-target flags, or poor transit-model fits. In particular, TOI-6075.01 has one of its two transits in Sector 75, outside the PM--EM1 baseline searched here, while TOI-5975.01 improves when EM2 is included but still fails our vetting criteria. Furthermore, \countess requires at least three transit observations to be considered a planet candidate, preventing the recovery of duo/mono transit TOIs. Restricting the search to PM and EM1 can therefore limit the recovery of longer-period TOIs, since additional sectors may provide extra transit events. However, the long-period TOIs missed in our sample, including EM2, did not provide the necessary additional transits for recovery.

We found that all 115 recovered TOIs have the same period as reported in the literature/ExoFOP catalog, but our radii seem to be systematically smaller compared to the catalog values (see Figure~\ref{fig:1to1P} and~\ref{fig:1to1}). The inconsistency in the planetary radius measurements is expected because the fitting methodology and stellar parameters adopted by \countess differ from those used by the individual pipelines that derived the published parameters for each TOI.  In addition, the planet and stellar radii reported in ExoFOP are heterogeneously derived; some incorporate follow-up ground-based observations, which can improve parameter precision relative to fits based solely on TESS photometry. We prioritize a uniform fitting regime over reproducing each TOI radius to construct a homogeneous planetary catalog to perform future demographic analyses. 

%Note that detrending is most efficient on a case-by-case basis, where you would vary the window length or detrending method based on each star.

\subsubsection{Multi-planet recovery}
There are 19 known TOI multi-planet systems within the TESS northern CVZ; 16 contain two transiting planets, and 3 contain three transiting planets. To conduct the multi-planet search, we first recovered the strongest signal by doing a \texttt{GERBLS} search. Once this signal passed vetting, we masked each transiting event within the combined pre-detrended light curves and reran the \texttt{GERBLS} search. For the triple systems, we conducted this masking an additional time. We successfully recovered 12 secondary candidates in multi-planet systems. \countess's inability to recover the tertiary and secondary TOIs in the multi-planet systems was also impacted by the low-S/N of the candidate signals. Furthermore, we applied this masked search to all TOIs known to host single-planet systems, but identified no new multi-planet candidates.

\subsection{Blind Search}

After testing \countess on known TOIs, we performed a blind transit search across the full sample of PM+EM1 stitched light curves of 26,114 FGKM stars in the TESS northern CVZ. This search was conducted without using any prior information from existing planet catalogs, enabling us to assess the pipeline’s ability to identify new transit signals in an unbiased manner.

We initially ran the targets through \texttt{GERBLS} to identify possible periodic signals. We searched over a 1--100 day period range with a window length of 0.5 days, then expanded to a 100--800 day period range with a window length of 1.75 days, as discussed in Section~\ref{sec:lcdetrend}. Of the 26,114 targets, we recovered a total of 5227 TCEs with an SNR $\geq$ 6. We then ran the \texttt{EXOTIC}-fitted blind-search TCEs through LEOVetter vetting to remove FAs and FPs, yielding 153 recovered planetary signals with 38 new signals. Each of these new signals was then run again through a multi-planet masked search to check for any missed companion planets in the systems, yielding two additional detections that passed vetting. The orbital period of these detections ranges from $<1$ day to $\sim$$124$ days with planetary radii spanning from $\sim$1$R_\oplus$ to $\sim$16$R_\oplus$. We then applied the pixel-level vetting test in \texttt{LEOVetter} to determine whether the signal originated from the target star. Of the 38 new detections, 24 have their signal detected off-center from the target star, leaving the 14 pixel-vetted targets to be labeled as fully vetted planet candidates. Finally, all new planet candidates were passed through \texttt{triceratops} to statistically validate or determine if they are FPs \citep{giacalone_vetting_2021}.

\begin{deluxetable*}{cccccccccc}[htb!]
\tabletypesize{\footnotesize}
\renewcommand{\arraystretch}{1.16}
\setlength{\tabcolsep}{3pt}
\tablecaption{New candidates discovered with \countess identified in this work with their corresponding \texttt{triceratops} FPP and NFPP values. Reported planet radii are derived from \texttt{EXOTIC} fits and include asymmetric uncertainties. Planets that pass the statistical validation thresholds from \texttt{triceratops} have been given a `b' designation.\label{tab:COIs}}

\tablehead{
\colhead{TIC ID} &
\colhead{$P$} &
\colhead{$R_p$} &
\colhead{$R_\star$} &
\colhead{$M_\star$} &
\colhead{$T_{\rm eff}$} &
\colhead{FPP} &
\colhead{NFPP} &
\colhead{SNR}
\vspace{-0.75em}\\ 
\colhead{} &
\colhead{(days)} &
\colhead{($R_\oplus$)} &
\colhead{($R_\odot$)} &
\colhead{($M_\odot$)} &
\colhead{(K)} &
\colhead{} &
\colhead{}
}

\startdata
219893931~b & $6.410389 \pm 0.000151$ & $2.92 \pm 0.37$ & $1.284 \pm 0.095$ & $1.140 \pm 0.092$ & $5956 \pm 174$ & $0.009 \pm 0.001$ & $(1.14 \pm 0.10)\times10^{-5}$ & 7.73\\
219893931.02 & $34.619938 \pm 0.000563$ & $4.19 \pm 0.38$ & $1.284 \pm 0.095$ & $1.140 \pm 0.092$ & $5959 \pm 175$ & $0.218 \pm 0.032$ & $0.004 \pm 0.000$ & 8.70 \\
237254473~b & $10.117702 \pm 0.000159$ & $3.30 \pm 0.29$ & $1.362	\pm 0.091$ & $1.120 \pm 0.090$ & $5700 \pm 167$ & $0.0139 \pm 0.001$ & $(8.99 \pm 0.61)\times 10^{-5}$ & 6.54\\
199632879.01 & $25.802505 \pm 0.000625$ & $2.84 \pm 0.28$ & $0.746 \pm 0.053$ & $0.688 \pm 0.055$ & $4640 \pm 136$ & $0.032 \pm 0.002$ & $0.002 \pm 0.000$ & 6.68 \\
353855214.01 & $23.169644 \pm 0.000558$ & $2.57 \pm 0.22$ & $1.131 \pm 0.070$ & $1.095 \pm 0.088$ & $6047 \pm 177$ & $0.041 \pm 0.003$ & $0.004 \pm 0.000$ & 7.31 \\
459982800.01 & $8.003264 \pm 0.000126$ & $2.28 \pm 0.24$ & $0.841 \pm 0.069$ & $0.855 \pm 0.069$ & $5435 \pm 159$ & $0.050 \pm 0.003$ & $0.022 \pm 0.001$ & 8.00 \\
219855901.01 & $1.198508 \pm 0.000019$ & $3.31 \pm 0.38$ & $1.042 \pm 0.075$ & $0.991 \pm 0.080$ & $5711 \pm 167$ & $0.073 \pm 0.001$ & $0.005 \pm 0.000$ & 15.31 \\
219822106.01 & $8.498852 \pm 0.000200$ & $2.82 \pm 0.29$ & $1.189 \pm 0.085$ & $1.082 \pm 0.088$ & $5840 \pm 171$ & $0.078 \pm 0.002$ & $0.026 \pm 0.001$ & 6.97 \\
219778735.01 & $7.457793 \pm 0.000120$ & $2.17 \pm 0.21$ & $0.845 \pm 0.056$ & $0.854 \pm 0.068$ & $5387 \pm 158$ & $0.099 \pm 0.002$ & $0.019 \pm 0.001$ & 7.94 \\
219812472.01 & $2.532999 \pm 0.000040$ & $1.73 \pm 0.46$ & $1.261 \pm 0.072$ & $1.087 \pm 0.088$ & $5728 \pm 168$ & $0.447 \pm 0.003$ & $0.026 \pm 0.000$ & 9.70 \\
229451673.01 & $29.358044 \pm 0.000485$ & $6.11 \pm 0.54$ & $1.481 \pm 0.103$ & $1.325 \pm 0.107$ & $6416 \pm 188$ & $0.594 \pm 0.051$ & $(5.73 \pm 0.53)\times10^{-7}$ & 14.72 \\
229451673.02 & $18.434541 \pm 0.000298$ & $5.91 \pm 0.53$ & $1.481 \pm 0.103$ & $1.325 \pm 0.107$ & $6416 \pm 188$ & $0.624 \pm 0.044$ & $0.012 \pm 0.001$ & 11.77 \\
219786666.01 & $15.182999 \pm 0.000252$ & $5.50 \pm 1.46$ & $0.951 \pm 0.063$ & $0.890 \pm 0.072$ & $5350 \pm 157$ & $0.718 \pm 0.013$ & $0.372 \pm 0.010$ & 7.08 \\
441741446.01 & $18.826798 \pm 0.000298$ & $11.10 \pm 0.78$ & $1.084 \pm 0.076$ & $0.840 \pm 0.068$ & $4748 \pm 139$ & $1.000 \pm 0.069$ & $0.672 \pm 0.251$ & 84.47 \\
\enddata

\tablecomments{FPP and NFPP are reported to three decimal places.}
\end{deluxetable*}

\subsection{Astrophysical False Positives}

Although our pixel-vetting procedure reduced our sample of 153 signals to a smaller dataset of 129 TOIs and new candidates, astrophysical false positives can remain in the final candidate list. %\texttt{triceratops} estimates the probability that a transit signal is caused by a planet relative to a suite of astrophysical false positive scenarios. 
To further assess the likelihood that FPs cause these signals, we evaluated all 14 new pixel-vetted candidates (see Table~\ref{tab:COIs}) 
%and 113 PCs, CPs, and KPs (see Table~\ref{tab:TOIs}) 
using the \texttt{triceratops} Bayesian validation framework.

%\texttt{triceratops} quantifies the probability that a given transit signal arises from a planet orbiting the target star relative to a list of astrophysical false-positive scenarios, including EBs associated with the target star, nearby stars, or unresolved background sources.
The results of the \texttt{triceratops} analysis are summarized in Table~\ref{tab:COIs}. Among the 14 new candidates that pass pixel vetting, eight have $\mathrm{FPP} < 0.5$ and nearby false-positive probabilities $\mathrm{NFPP} < 10^{-1}$, and two (TIC 237254473.01 and TIC 219893931.01) have $\mathrm{FPP} < 0.015$ and $\mathrm{NFPP} < 10^{-3}$, classifying them into different tiers of planetary likelihood under the criteria outlined in \cite{giacalone_triceratops_2020} \citep[see more detail in][]{giacalone_vetting_2021}. The remaining four candidates are dominated by FP scenarios, most commonly involving nearby EBs (NFPP values $>10^{-1}$), consistent with the outcomes of the pixel-level vetting.

\subsection{New Planet Candidates}

\begin{figure*}[t]
    \centering
    \includegraphics[width=0.85\textwidth]{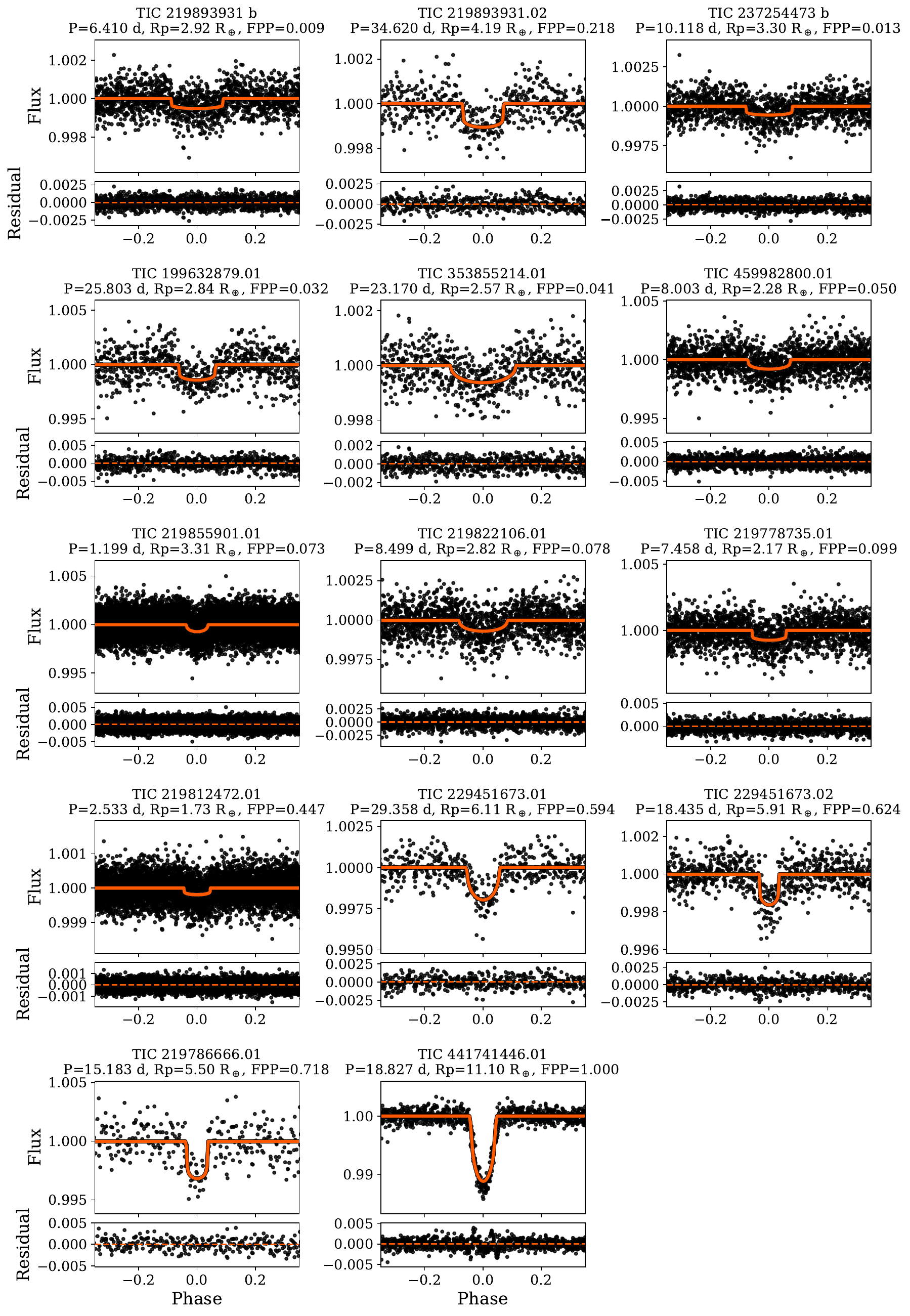}
    \caption{The phase-folded transits of the new planet candidates discovered with \countess. We list the planetary and \texttt{triceratops} parameters in Table~\ref{tab:COIs}.}
    \label{fig:COIsLCs}
\end{figure*}

\begin{figure*}[t]
    \centering
    \includegraphics[width=1.05\textwidth]{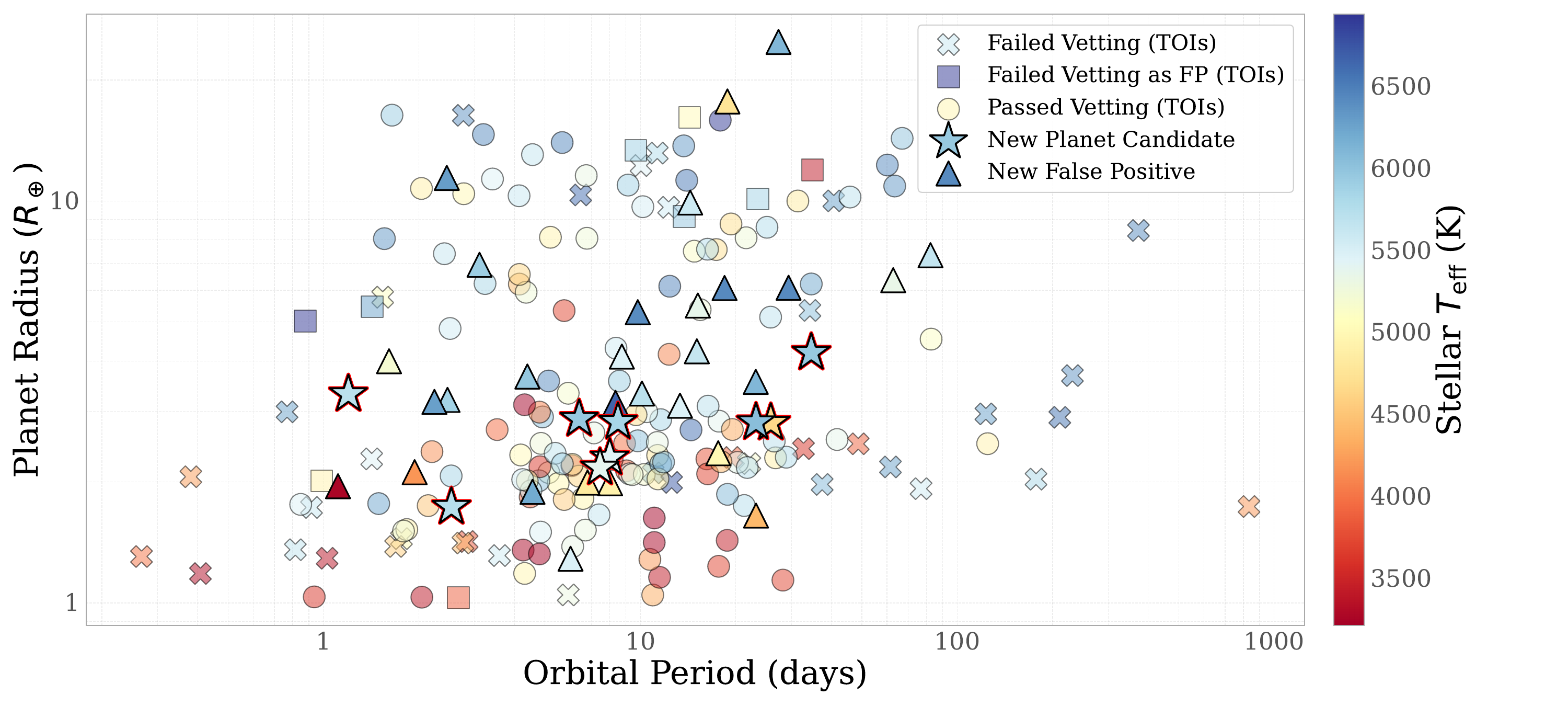}
    \caption{The distribution of planetary radius and orbital period of TOIs and new false positives and planet candidates. \countess discovered 38 new planetary signals, with 28 of them being FPs through \texttt{triceratops} validation and \texttt{LEOVetter} pixel-vetting. Note that the square symbols indicate TOI FPs that were successfully recovered as FPs through \countess.}
    \label{fig:RadvPer}
\end{figure*}

After we ran statistical validation on the 14 new vetted candidates, 10 were classified as planet candidates by the \texttt{triceratops} validation. These systems span orbital periods from a few days to several tens of days (see Figure~\ref{fig:COIsLCs}). TIC 219893931.02 has the largest orbital period and planet radius of the new candidates, at slightly larger than Neptune's radius ($P  \approx 34.62$ days; $R_p = 4.19R_\oplus$). Although its $\mathrm{FPP} < 0.5$ and $\mathrm{NFPP} < 10^{-1}$, this signal represents a detection at $>3\times$ the maximum orbital period that can be robustly identified within a single $\sim$27-day TESS sector. All other new ``likely planet'' candidates fall within the sub-Neptune category of planets, but two additional candidates, TIC 199632879.01 ($P  \approx 25.80$; $R_p = 2.84R_\oplus$) and TIC 353855214.01 ($P  \approx 23.17$; $R_p = 2.57R_\oplus$), have orbital periods $\gtrsim$ 10-days. Figure~\ref{fig:RadvPer} details the period and planetary radii distribution of recovered TOIs and new candidates. In addition to the primary signals for these targets, we searched for secondary and tertiary signals. The multi-planet search yielded two instances of secondary signals for TIC 219893931 and TIC 229451673.

Furthermore, TIC 219893931.01 having an $\mathrm{FPP} < 0.015$ and $\mathrm{NFPP} < 10^{-3}$. This planet was initially discovered as a community planet candidate by \cite{montalto_search_2023} with a $P  \approx$ 6.41 and $R_p = 3.48 \pm 1.27$. TIC 237254473.01 ($P  \approx 10.12$; $R_p = 3.30R_\oplus$) also has an $\mathrm{FPP} < 0.015$ and $\mathrm{NFPP} < 10^{-3}$. Both of these planet candidates fall within the \texttt{triceratops} criteria of a statistically validated planet, so we designate them TIC~219893931~b and TIC~237254473~b. 

We also estimated the expected radial-velocity semi-amplitudes and atmospheric follow-up metrics for the two statistically validated candidates. Specifically, we computed the transmission spectroscopy metric and emission spectroscopy metric \citep[TSM/ESM;][]{kempton_framework_2018}. Because neither planet currently has a measured mass, we estimated $M_p$ using the mass--radius relations of \citet{chen_probabilistic_2016}; these values should therefore be interpreted as predicted follow-up metrics rather than measured system properties. For TIC~219893931~b, we estimate $M_p \approx 11.96~M_\oplus$, yielding $\mathrm{TSM}=20.47$ and $\mathrm{ESM}=1.90$. For TIC~237254473~b, we estimate $M_p \approx 10.93~M_\oplus$, yielding $\mathrm{TSM}=14.26$ and $\mathrm{ESM}=1.13$. Assuming circular, edge-on orbits, the corresponding predicted radial-velocity semi-amplitudes are $K \approx 3.77~\mathrm{m~s^{-1}}$ and $K \approx 3.00~\mathrm{m~s^{-1}}$, respectively.
%TIC 237254473.01 ($P  \approx 10.12$; $R_p = 3.30R_\oplus$) is a sub-Neptune right on the cusp between a hot ($P  < 10$) and warm Neptune ($P  > 10$), and TIC 219893931.02 is a clear hot sub-Neptune.
All of the new planet candidates orbit FGK stars, with a majority of them being G-dwarf stars, and are labeled with their validation results in Table~\ref{tab:COIs}. 

%Finally, we emphasize that this proof-of-concept search represents only a fraction of the stars available in the TESS northern CVZ. There are $\gtrsim14\times$ more TESS northern CVZ targets that can be processed with TESS FFIs than the subset analyzed here. As the pipeline is scaled to the full northern -- and, in the future, southern -- CVZ stellar sample, \countess has the potential to yield a substantially larger set of vetted candidates pushing TESS coverage out to $>$100 days.

\section{Summary \& Future work}\label{sec:Summary}

The TESS CVZs enable long-baseline observations of transiting exoplanets around nearby stars with TESS, rivaling the orbital period coverage achieved by \kepler and \ktwo. In this study, we aimed to robustly identify transiting planets that are challenging to recover with standard short-baseline single-sector searches, either due to long orbital periods or low signal-to-noise ratios. To achieve this,
%we developed \countess, a modular and automated transit-search pipeline that combines light curves across multiple TESS sectors and cadences, incorporates transit duration variable detrending, and performs robust vetting and validation. To summarize,
\begin{itemize}
    \item We constructed a homogeneously characterized stellar sample of 391,059 FGKM stars in the TESS northern CVZ using the same stellar parameter derivation framework as \citet{Hardegree-Ullman2025}. This approach ensures that stellar effective temperatures, radii, masses, and metallicities are derived uniformly and places the TESS northern CVZ stellar sample on the same footing as the \kepler and \ktwo datasets, thereby enabling homogeneous derivation of planetary properties and therefore, robust exoplanet demographics across surveys.

    \item We developed \countess, a modular transit-search pipeline designed to perform homogeneous long-baseline searches across multiple TESS sectors and cadences. By combining PM and EM1 FFI light curves, \countess enables consistent biweight detrending from \texttt{W\=otan}, \texttt{GERBLS} fast-folding BLS period searching, \texttt{LEOVetter} vetting, and \texttt{triceratops} statistical validation.

    \item We trained and tested \countess by conducting a recovery search for known TOIs in the TESS northern CVZ. We were able to recover 115 out of the 159 PC, KP, and CP transit signals (72\%), 12 of which were secondary signals apart from multi-transiting planet systems. We found that the TOIs that we were not able to recover were due to either gaps in TESS PM or EM1 observations or those significantly affected by noise. All 115 recovered TOIs have periods consistent with the literature or ExoFOP catalog values; however, our recovered planet radii are systematically smaller due to differences in fitting methods and adopted stellar parameters.

    \item We conducted a blind, systematic search for transiting planets on the light curves of 26,114 FGKM stars in the TESS northern CVZ. We identified 14 new candidates that passed the \texttt{LEOVetter} vetting (Table~\ref{tab:COIs}), and 10 that passed the  \texttt{triceratops} ``likely planets'' thresholds of $\mathrm{FPP} < 0.5$ and $\mathrm{NFPP} < 10^{-1}$.

    \item We discovered two new sub-Neptunes, TIC~219893931~b ($P = 6.41$ days; $R_p=2.92\ R_\oplus$) and TIC~237254473~b ($P = 10.12$ days; $R_p=3.3\ R_\oplus$) that passed the \texttt{triceratops} statistical validation thresholds of $\mathrm{FPP} < 0.015$ and $\mathrm{NFPP} < 10^{-3}$.
\end{itemize}

Our search of the TESS northern CVZ with \countess demonstrates that TESS can be effectively used to perform long-baseline transit searches across a variety of cadences and TESS sectors.
% ($P$ = 0.85--124 days) and radii ($R_\text{p}$ = 1--16$R_\oplus$) that overlap with a regime previously dominated by \kepler and \ktwo discoveries. 
%In total, we recovered 113 and discovered 14 new candidates that were then passed through to \texttt{triceratops}, where we use their phase-folded light curves to validate the planetary signals that were separate from the astrophysical false positives. 
Our sample of 10 new planet candidates and 2 new statistically validated planets highlights the potential of \countess to discover transiting planets with $P > 10$ days, expanding TESS's opportunities to better understand planet populations in the solar neighborhood. Importantly, this proof-of-concept search represents only a small fraction of the available CVZ stellar sample: there are $\gtrsim14\times$ more TESS northern CVZ targets accessible through other TESS FFIs than were analyzed here with TESS-SPOC. In future work, we will scale the pipeline to the full TESS northern CVZ and extend it to the TESS southern CVZ, enabling \countess to produce a substantially larger vetted candidate sample and push TESS sensitivity to orbital periods beyond 100 days. Additionally, a future feature of \countess would be to incorporate true and false inclusion probabilities \citep[TIP/FIP;][]{hara_detecting_2022}. TIP/FIP provides a Bayesian model-comparison to assess whether a planet signal should be included within a specified region of period parameter space. Such an approach would be particularly valuable for long-period candidates with only a few observed transits, where individual events may be difficult to distinguish from systematics or isolated light-curve features.

The European Space Agency PLAnetary Transits and Oscillations \citep[PLATO;][]{rauer_plato_2025, cabrera_assessment_2026} mission, expected to launch in early 2027, will continue the search for new planets and aid in the monitoring and follow-up observations of currently known candidate and confirmed exoplanets. PLATO's footprint will cover 2132 square degrees, and the current plan is to observe two separate regions for two years each, called Long-duration Observation Phase (LOP) fields; LOPS2 for the first two years in the southern hemisphere, and will potentially move to observe the proposed LOPN1 for the following two years in the northern hemisphere \citep{Nascimbeni2022, Eschen2024, Nascimbeni2025}. The LOPN1 field overlaps entirely with the original Kepler field and will cover up to about half of the TESS northern CVZ, whereas the LOPS2 field overlaps with about 2/3 of the TESS southern CVZ \citep{Eschen2024, Nascimbeni2025}. PLATO will be able to continue monitoring TESS CVZ targets, and its greater sensitivity to smaller, longer-period planets increases the chances of finding new long-period Earth-sized exoplanets \citep{Eschen2024}.

Our newly discovered exoplanet candidates underscore the scientific value of continued exploration of the TESS CVZs and highlight the mission’s promising capability to perform demographic studies that build upon the discoveries of \kepler and \ktwo. A uniform, end-to-end pipeline such as \countess is essential for enabling robust occurrence rate and population-level analyses. While \kepler and \ktwo revealed fundamental features of exoplanet populations, including the radius valley and hot Neptune desert, those surveys primarily probed distant stellar populations. By exploiting the extended temporal baselines of the TESS CVZs, \countess will enable these demographic features to be investigated in the solar neighborhood for the first time. The CVZs also provide a unique opportunity to extend these analyses to long-period planets around M dwarfs, a stellar population underrepresented in \kepler and \ktwo. 
%With a $\sim$600--1000\,nm bandpass, TESS is particularly well suited to observe nearby M dwarfs, whose spectral energy distributions peak in the near-infrared ($\sim$1000\,nm). 
The planets discovered around these stars will be prime targets for spectroscopic follow-up, and detailed constraints on their compositions as a function of orbital period would provide valuable insight into small-planet formation and M dwarf planetary atmospheres. Together, these efforts position \countess to transform the TESS CVZs into a powerful laboratory for long-period exoplanet demographics in the solar neighborhood.

\section{Acknowledgments}
%thank HPF Check what is needed
AH and SM acknowledge support from NSF grant AST-2108512. MK acknowledges the support of the Natural Sciences and Engineering Research Council of Canada (NSERC), RGPIN-2024-06452. Cette recherche a été financée par le Conseil de recherches en sciences naturelles et en génie du Canada (CRSNG), RGPIN-2024-06452.

% PSU Land acknowledgement
The Pennsylvania State University campuses are located on the original homelands of the Erie, Haudenosaunee (Seneca, Cayuga, Onondaga, Oneida, Mohawk, and Tuscarora), Lenape (Delaware Nation, Delaware Tribe, Stockbridge-Munsee), Shawnee (Absentee, Eastern, and Oklahoma), Susquehannock, and Wahzhazhe (Osage) Nations.  As a land grant institution, we acknowledge and honor the traditional caretakers of these lands and strive to understand and model their responsible stewardship. We also acknowledge the longer history of these lands and our place in that history. Computations for this research were performed on the Pennsylvania State University’s Institute for Computational
and Data Sciences Advanced CyberInfrastructure (ICDS-ACI). This content is solely the responsibility of the authors and does not necessarily represent the views of the Institute for Computational and Data Sciences. The Center for Exoplanets and Habitable Worlds is supported by the Pennsylvania State University, the Eberly College of Science, and the Pennsylvania Space Grant Consortium.

%thank HiPerGator and other UF things
We acknowledge that for thousands of years the area now comprising the state of Florida has been, and continues to be, home to many Native Nations. We further recognize that the main campus of the University of Florida is located on the ancestral territory of the Potano and of the Seminole peoples. The Potano, of Timucua affiliation, lived here in the Alachua region from before European arrival until the destruction of their towns in the early 1700s. The Seminole, also known as the Alachua Seminole, established towns here shortly after but were
forced from the land as a result of a series of wars with the United States known as the Seminole Wars. We, the authors, acknowledge our obligation to honor the past, present, and future Native residents and cultures
of Florida. The authors acknowledge the University of Florida Research Computing for providing computational resources and support that have contributed to the research results reported in this publication

% The Hobby–Eberly Telescope is a joint project of the University of Texas at Austin, the Pennsylvania State University, LudwigMaximilians-Universität München, and Georg-August Universität Gottingen. The HET is named in honor of its principal benefactors, William P. Hobby and Robert E. Eberly. The HET collaboration acknowledges the support and resources from the Texas Advanced Computing Center. We would like to acknowledge that the HET is built on Indigenous land. Moreover, we would like to acknowledge and pay our respects to the Carrizo and Comecrudo, Coahuiltecan, Caddo, Tonkawa, Comanche, Lipan Apache, Alabama–Coushatta, Kickapoo, Tigua Pueblo, and all the American Indian and Indigenous Peoples and communities who have been or have become a part of these lands and territories in Texas, here on Turtle Island.

This work has made use of data from the European Space Agency (ESA) mission \gaia (https://www.cosmos.esa.int/
gaia), processed by the \gaia Data Processing and Analysis Consortium (DPAC, https://www.cosmos.esa.int/web/gaia/
dpac/consortium). Funding for the DPAC has been provided by national institutions, in particular the institutions participating in the \gaia Multilateral Agreement. Some of the data presented in this paper were obtained from MAST at STScI. Support for MAST for non-HST data is provided by the NASA Office of Space Science via grant NNX09AF08G and by other grants and contracts. This work includes data collected by the TESS mission, which are publicly available from MAST \citep{tess_team_2021_lc,tess_team_2021_tpf}. Funding for the TESS mission is provided by the NASA Science Mission directorate. This research made use of the (i) NASA Exoplanet Archive, which is operated by Caltech, under contract with NASA under the Exoplanet Exploration Program, (ii) SIMBAD database, operated at CDS, Strasbourg, France, (iii) NASA’s Astrophysics Data System Bibliographic Services, and (iv) data from 2MASS, a joint project of the University of Massachusetts and IPAC at Caltech, funded by NASA and the NSF. This research has made use of the Exoplanet Follow-up Observation Program \citep[ExoFOP;][]{nexsci_exofop_2022} website, which is operated by the California Institute of Technology, under contract with the National Aeronautics and Space Administration under the Exoplanet Exploration Program.

\facilities{\gaia{}, TESS, Exoplanet Archive}
\software{
%\texttt{ArviZ} \citep{kumar_arviz_2019}, 
%AstroImageJ \citep{collins_astroimagej_2017}, 
% \texttt{astrometry.net} \citep{hogg_automated_2008},
% \texttt{astroquery} \citep{ginsburg_astroquery_2019}, 
\texttt{astropy} \citep{robitaille_astropy_2013, astropy_collaboration_astropy_2018},
%BANYAN \citep{gagne_banyan_2018},
%BANZAI \citep{mccully_real-time_2018},
%\texttt{barycorrpy} \citep{kanodia_python_2018}, 
\texttt{batman} \citep{kreidberg_batman_2015},
%\texttt{celerite2} \citep{foreman-mackey_fast_2017, foreman-mackey_scalable_2018},
% \texttt{DEATHSTAR} \citep{ross_deathstar_2023},
%\texttt{eleanor} \citep{feinstein_eleanor_2019},
% \texttt{EVEREST} \citep{luger_everest_2016, luger_update_2018},
%\texttt{EXOFASTv2} \citep{eastman_exofastv2_2019},
\texttt{exoplanet} \citep{foreman-mackey_exoplanet-devexoplanet_2021, foreman-mackey_exoplanet_2021},
\texttt{EXOTIC} \citep{zellem_utilizing_2020},
%\texttt{HPF-SpecMatch} \citep{stefansson_sub-neptune-sized_2020},
%\texttt{HxRGproc} \citep{ninan_habitable-zone_2018},
% \texttt{GALPY} \citep{bovy_galpy_2015},
\texttt{GERBLS} (Ment et al. subm.),
\texttt{ipython} \citep{perez_ipython_2007},
\texttt{LEOVetter} \citep{kunimoto_leo-vetter_2025},
\texttt{lightkurve} \citep{lightkurve_collaboration_lightkurve_2018},
\texttt{matplotlib} \citep{hunter_matplotlib_2007},
% \texttt{MOLUSC} \citep{wood_characterizing_2021},
% \texttt{MRExo} \citep{kanodia_mass-radius_2019},
\texttt{numpy} \citep{harris2020array},
\texttt{pandas} \citep{mckinney-proc-scipy-2010},
%\texttt{photutils} \citep{larry_bradley_2020},
%\texttt{pyastrotools} \citep{kanodia_2023},
\texttt{PyMC3} \citep{salvatier_probabilistic_2016},
\texttt{scipy} \citep{ virtanen_scipy_2020},
%\texttt{SERVAL} \citep{zechmeister_spectrum_2018},
%\texttt{tglc} \citep{han_tess-gaia_2023},
%\texttt{Theano} \citep{the_theano_development_team_theano_2016},
\texttt{TRICERATOPS} \citep{giacalone_vetting_2021},
% \texttt{VESPA} \citep{2015ascl.soft03011M, morton_false_2016}.
\texttt{W\={o}tan} \citep{hippke_wotan_2019}.
}

\clearpage

\appendix
\section{Skye Excess Metric}\label{app:sem}

\begin{figure*}[h]
    \centering
    \includegraphics[width=0.95\textwidth]{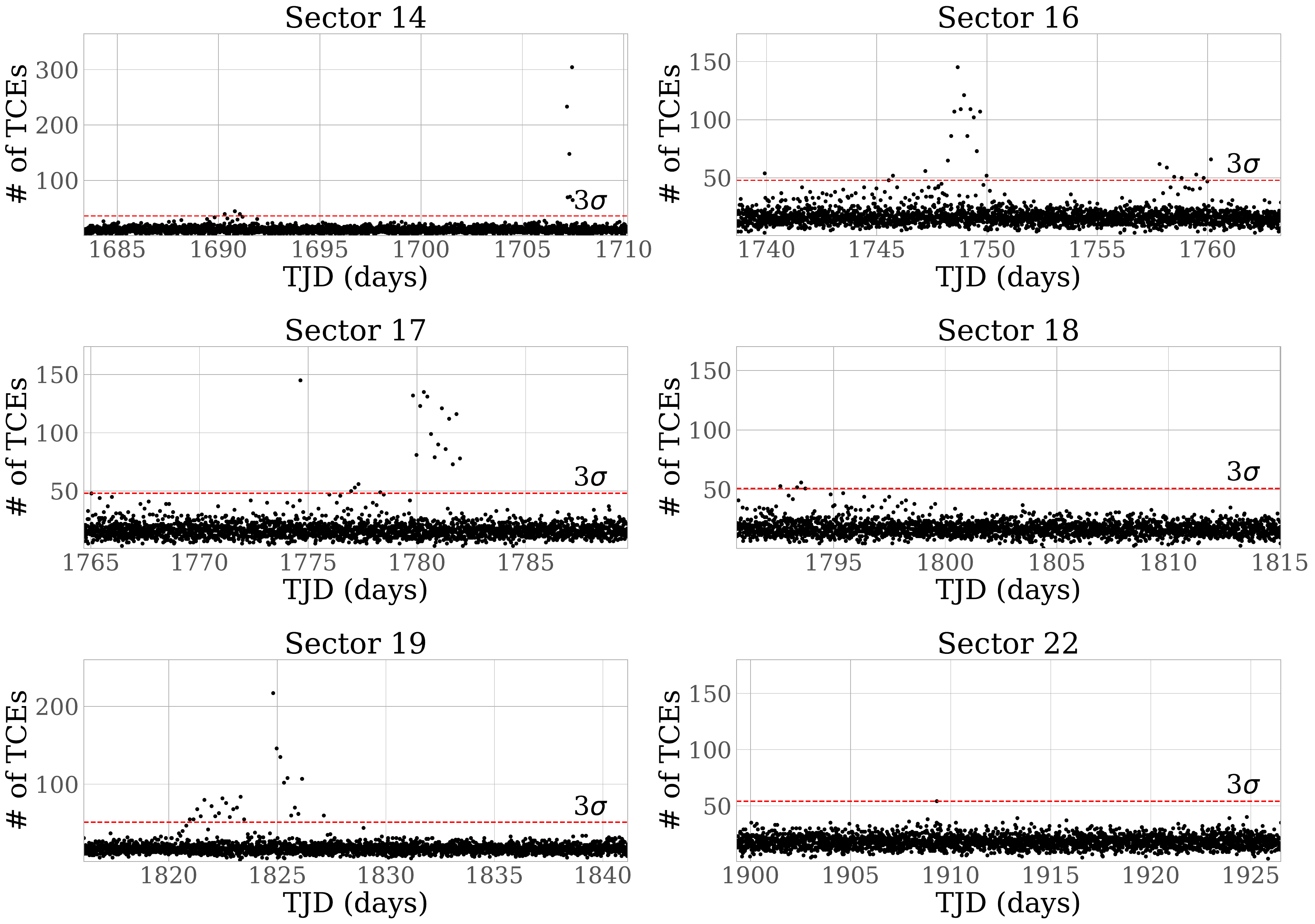}
    \caption{The Skye Excess Metric plots depicting the number of transiting events for each timestamp. The red dashed line is the 3$\sigma$ threshold.}
    \label{fig:SME}
\end{figure*}

Figure~\ref{fig:SME} shows the Skye Excess Metric (SEM) for all flagged sectors in our blind search, which we outlined in Section~\ref{sec:system}. All timestamps above the corresponding sector's 3$\sigma$ threshold were removed from the overall search.

\clearpage

\section{Window Length Fix}\label{app:WLF}

\begin{figure*}[h]
    \centering
    \includegraphics[width=\textwidth]{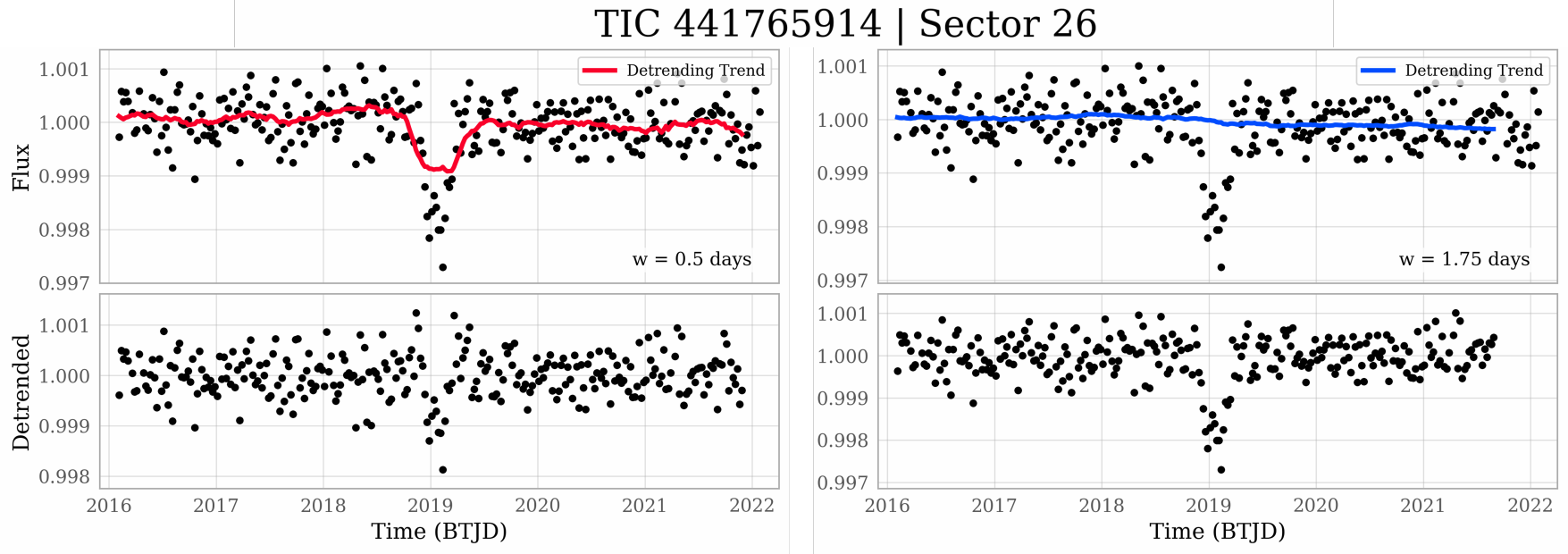}
    \caption{The comparison of different detrending window lengths on TOI-2088b ($P = 124.72$ days). The left and right plot show the detrending with a window length of 0.5 and 1.75 days respectively.}
    \label{fig:DetrendComp}
\end{figure*}

Figure~\ref{fig:DetrendComp} shows the comparison between using a short window length ($w = 0.5$ days) versus a longer window length ($w = 1.75$ days) to detrend a light curve with a transit that has a long duration. If we use a window length where $w \sim t_{\text{dur}}$, the detrending will remove a portion of the tranist depth and duration. Therefore we use a longer window length to preserve the full depth and duration of a long duration transit. 

\clearpage

\section{vetting test}\label{app:APPVet}

Table~\ref{tab:LEOThresh} outlines the list of \texttt{LEOVetter} false alarm and false positive test thresholds. If the target satisfies the threshold, then it fails that test and is classified as either an FA or an FP. 

\begin{table*}[ht]
\centering
\caption{False-alarm (FA) and astrophysical false-positive (FP) checks and their implemented threshold logic. A candidate is flagged if the listed condition evaluates to true.}
\label{tab:LEOThresh}
\renewcommand{\arraystretch}{1.15}
\begin{tabular}{llll}
\hline

\textbf{Case} & \textbf{Test label} & \textbf{Threshold(s)} & \textbf{Citation} \\
\hline
\hline
\multicolumn{3}{l}{FA}\\
\hline

 & Signal too weak & MES $< 6.2$ & \cite{kunimoto_leo-vetter_2025}\\
 & Not enough valid transits & MES $< 6.2$, $N_{\rm transit} < 3$ & \cite{thompson_planetary_2018} \\
 & Bad transit shape & SHP $> 0.5$ & \cite{coughlin_tcert_2017} \\
 & Non-unique events (global) & MS1, MS2, MS3 $< 1$ & \cite{coughlin_tcert_2017} \\
 & Non-unique events (local) & CHASES $< 0.78$ (if $N_{\rm transit} \le 5$) & \cite{thompson_planetary_2018} \\
 & Inconsistent depths & $0.5 < \mathrm{DMM} < 1.5$ & \cite{coughlin_tcert_2017} \\
 & Single-event dominated & max SES / MES $> 0.8$ & \cite{thompson_planetary_2018} \\
 & Poor transit fit & $\Delta$AIC $> -60$ or $-30$ & \cite{kunimoto_leo-vetter_2025} \\
 & Sinusoidal variability & SWEET $> 15$, $P < 10$ d & \cite{thompson_planetary_2018}\\
 & Unphysical duration & $a/R_\star < 1.5$ or $q_{\rm tran} > 0.5$ & \cite{kunimoto_leo-vetter_2025} \\
 & Asymmetric transit & ASYM $> 8$ & \cite{eschen_nine_2024} \\
 & Inconsistent SNRs & CHI $< 7.8$ & \cite{kunimoto_leo-vetter_2025}\\
 & Data gaps & Gap fraction $\ge 0.5$ & \cite{kunimoto_leo-vetter_2025}\\
\hline
\multicolumn{3}{l}{FP}\\
\hline
 & Odd--even differences & Depth or epoch significance $> 3$--$10\sigma$ & \cite{kunimoto_leo-vetter_2025}\\
 & V-shaped transit & $b + R_p/R_\star > 1.5$ & \cite{thompson_planetary_2018}\\
 & Object too large & $R_p > 22\,R_\oplus$ & \cite{kunimoto_leo-vetter_2025}\\
 & Significant secondary & MS4 $> 0$, MS5/MS6 $> -1$ & \cite{kunimoto_leo-vetter_2025}\\
 & Off-target source & Offset quality $> 15$ & \cite{bryson_identification_2013}\\
\hline
\end{tabular}
\end{table*}

\clearpage

\section{\texttt{EXOTIC} Thresholds}\label{app:exThresh}

\begin{table}[h!]
\centering
\caption{Search parameter bounds adopted for transit fitting.}
\label{tab:thresh}
\renewcommand{\arraystretch}{1.25}
\begin{tabular}{|c|c|c|}
\hline
\textbf{Parameter} & \textbf{Threshold Range} & \textbf{Notes} \\
\hline

$R_p/R_\star$ &
$0 \le R_p/R_\star \le 0.20/R_\star$ &
$R_p \lesssim 22\,R_\oplus$ \\[4pt]
\hline

$a/R_\star$ &
$\displaystyle
\frac{\left[M_\star(P_{\min}/365.25)^2\right]^{1/3}}{R_\star}
\le
\frac{a}{R_\star}
\le
\frac{\left[M_\star(P_{\max}/365.25)^2\right]^{1/3}}{R_\star}
$ &
$\sim$0.008--0.669 AU (solar-type star) \\[6pt]
\hline

$i$ &
$\displaystyle
\arccos\!\left(
\frac{R_\star (6.957\times10^{10})}
{\left[M_\star (P_{\min}/365.25)^2\right]^{1/3}(1.496\times10^{13})}
\right)
\le i \le 90^\circ
$ &
$b \lesssim 1$ \\[4pt]
\hline

\shortstack{$u_0$ \\[6pt] $u_1$} &
\shortstack{\\[6pt]$0.2 \le u_0 \le 0.6$ \\[6pt] $0.2 \le u_1 \le 0.6$} &
\raisebox{9pt}{Quadratic limb-darkening coefficients} \\[6pt]
\hline

\end{tabular}
\end{table}

\clearpage

\section{recovered tois}\label{app:TOIs}

Table~\ref{tab:TOIs} compares the orbital periods and planet radii recovered by COUNTESS with those published in ExoFOP. These TOIs range from hot Earths to warm Jupiters.

\begin{center}
\scriptsize
\setlength{\tabcolsep}{6pt}
\renewcommand{\arraystretch}{1.15}

\begin{longtable}{ccccccc}
\caption{Recovered vs.\ ExoFOP orbital periods and planet radii for candidate planets.}
\label{tab:TOIs}\\
\toprule
TIC ID & TOI ID &
$P $ (rec; days) &
$P $ (TOI; days) &
$R_p$ (rec; $R_\oplus$) &
$R_p$ (TOI; $R_\oplus$) &
TFOP Disposition \\
\midrule
\endfirsthead

\toprule
TIC ID & TOI ID &
$P $ (rec; days) &
$P $ (pub; days) &
$R_p$ (rec; $R_\oplus$) &
$R_p$ (pub; $R_\oplus$) &
TFOP Disposition \\
\midrule
\endhead

\midrule
\multicolumn{7}{r}{\textit{Continued on next page}}\\
\endfoot

\bottomrule
\endlastfoot

\multicolumn{7}{c}{\textbf{Recovered TOIs}}\\
\midrule
233535738 & 4106.01 & $16.3563 \pm 0.0003$ & $16.3559 \pm 7.66e-05$ & $3.09 \pm 0.25$ & $3.48 \pm 0.21$ & APC \\
232641326 & 1874.01 & $17.8683 \pm 0.0003$ & $17.8684 \pm 3.52e-05$ & $15.89 \pm 1.14$ & $19.38 \pm 0.84$ & APC \\
356978132 & 1755.01 & $60.1855 \pm 0.0010$ & $60.1850 \pm 8.31e-05$ & $12.29 \pm 0.85$ & $11.00 \pm 1.65$ & APC \\
229747848 & 1347.01 & $0.8474 \pm 1.34e-05$ & $0.8474 \pm 6.10e-07$ & $1.76 \pm 0.20$ & $1.81 \pm 0.10$ & CP\textsuperscript{a} \\
287139872 & 1752.01 & $0.9352 \pm 1.49e-05$ & $0.9352 \pm 2.00e-06$ & $1.04 \pm 2.51$ & $1.69 \pm 0.07$ & CP\textsuperscript{b} \\
377293776 & 1450.01 & $2.0440 \pm 0.0074$ & $2.0439 \pm 1.00e-06$ & $1.03 \pm 0.06$ & $1.13 \pm 0.04$ & CP\textsuperscript{c} \\
390021939 & 4030.01 & $2.5092 \pm 0.0003$ & $2.5090 \pm 3.51e-05$ & $4.82 \pm 0.35$ & $5.85 \pm 0.42$ & CP\textsuperscript{d} \\
219852584 & 1295.01 & $3.1968 \pm 5.00e-05$ & $3.1969 \pm 5.00e-07$ & $14.64 \pm 0.91$ & $15.69 \pm 0.90$ & CP\textsuperscript{e} \\
229400092 & 1826.01 & $4.1421 \pm 6.48e-05$ & $4.1420 \pm 1.20e-06$ & $10.31 \pm 0.76$ & $12.59 \pm 0.15$ & CP\textsuperscript{f} \\
198241702 & 1269.01 & $4.2531 \pm 6.63e-05$ & $4.2530 \pm 6.80e-06$ & $2.03 \pm 0.17$ & $2.40 \pm 0.15$ & CP\textsuperscript{g} \\
219860288 & 1743.01 & $4.2660 \pm 6.82e-05$ & $4.2660 \pm 2.00e-06$ & $1.35 \pm 0.11$ & $1.83 \pm 0.11$ & CP\textsuperscript{h} \\
198390247 & 1453.02 & $4.3135 \pm 6.86e-05$ & $4.3135 \pm 1.34e-05$ & $1.18 \pm 0.12$ & $1.13 \pm 0.09$ & CP\textsuperscript{i} \\
232612416 & 1248.01 & $4.3602 \pm 6.83e-05$ & $4.3602 \pm 1.10e-06$ & $5.93 \pm 0.46$ & $6.81 \pm 0.12$ & CP\textsuperscript{g} \\
233602827 & 1749.01 & $4.4892 \pm 7.00e-05$ & $4.4891 \pm 8.23e-06$ & $1.84 \pm 0.12$ & $2.00 \pm 1.01$ & CP\textsuperscript{j} \\
259168516 & 1680.01 & $4.8026 \pm 7.51e-05$ & $4.8027 \pm 5.32e-06$ & $1.32 \pm 0.09$ & $1.42 \pm 0.30$ & CP\textsuperscript{k} \\
229747848 & 1347.02 & $4.8420 \pm 7.58e-05$ & $4.8419 \pm 1.20e-05$ & $1.50 \pm 0.15$ & $1.68 \pm 0.10$ & CP\textsuperscript{a} \\
229650439 & 1438.02 & $5.1398 \pm 8.05e-05$ & $5.1397 \pm 3.00e-06$ & $2.11 \pm 0.17$ & $3.04 \pm 0.19$ & CP\textsuperscript{l} \\
289580577 & 1753.01 & $5.3847 \pm 8.43e-05$ & $5.3846 \pm 9.90e-06$ & $2.36 \pm 0.18$ & $2.48 \pm 0.10$ & CP\textsuperscript{g} \\
219852882 & 1346.02 & $5.5026 \pm 8.60e-05$ & $5.5026 \pm 1.50e-05$ & $1.98 \pm 0.16$ & $2.23 \pm 0.14$ & CP\textsuperscript{m} \\
198360694 & 2071.01 & $5.6739 \pm 8.90e-05$ & $5.6740 \pm 1.02e-05$ & $2.22 \pm 0.20$ & $2.56 \pm 0.91$ & CP\textsuperscript{d} \\
233087860 & 1184.01 & $5.7485 \pm 9.04e-05$ & $5.7484 \pm 3.90e-06$ & $1.81 \pm 0.13$ & $2.41 \pm 0.14$ & CP\textsuperscript{g} \\
441738827 & 2084.01 & $6.0785 \pm 9.97e-05$ & $6.0784 \pm 6.58e-05$ & $2.20 \pm 0.16$ & $2.07 \pm 0.17$ & CP\textsuperscript{n} \\
356867115 & 1301.01 & $6.0965 \pm 9.55e-05$ & $6.0964 \pm 3.60e-06$ & $2.21 \pm 0.19$ & $2.48 \pm 0.20$ & CP\textsuperscript{o} \\
219850915 & 1244.01 & $6.4003 \pm 0.0001$ & $6.4003 \pm 1.30e-05$ & $2.07 \pm 0.18$ & $2.38 \pm 0.13$ & CP\textsuperscript{g} \\
198390247 & 1453.01 & $6.5888 \pm 0.0001$ & $6.5887 \pm 4.80e-06$ & $1.82 \pm 0.14$ & $2.14 \pm 0.15$ & CP\textsuperscript{i} \\
233602827 & 1749.02 & $9.0446 \pm 0.0001$ & $9.0447 \pm 1.67e-05$ & $2.13 \pm 0.12$ & $2.25 \pm 1.36$ & CP\textsuperscript{j} \\
198241702 & 1269.02 & $9.2379 \pm 0.0002$ & $9.2379 \pm 4.01e-05$ & $2.10 \pm 0.17$ & $2.23 \pm 0.24$ & CP\textsuperscript{d} \\
229650439 & 1438.01 & $9.4282 \pm 0.0001$ & $9.4281 \pm 1.00e-06$ & $2.09 \pm 0.16$ & $2.75 \pm 0.14$ & CP\textsuperscript{l} \\
237222864 & 1255.01 & $10.2890 \pm 0.0002$ & $10.2889 \pm 1.02e-05$ & $2.09 \pm 0.13$ & $2.73 \pm 0.17$ & CP\textsuperscript{p} \\
420112589 & 1452.01 & $11.0620 \pm 0.0002$ & $11.0620 \pm 1.53e-05$ & $1.41 \pm 0.07$ & $1.62 \pm 0.22$ & CP\textsuperscript{q} \\
237099296 & 1750.01 & $11.3371 \pm 0.0002$ & $11.3373 \pm 4.22e-05$ & $2.33 \pm 0.18$ & $2.79 \pm 0.34$ & CP\textsuperscript{d} \\
235678745 & 2095.01 & $17.6653 \pm 0.0003$ & $17.6649 \pm 4.80e-05$ & $1.23 \pm 0.07$ & $1.30 \pm 0.49$ & CP\textsuperscript{r} \\
198356533 & 1437.01 & $18.8387 \pm 0.0003$ & $18.8409 \pm 6.80e-05$ & $1.86 \pm 0.15$ & $2.24 \pm 0.23$ & CP\textsuperscript{s} \\
219857012 & 1742.01 & $21.2681 \pm 0.0003$ & $21.2691 \pm 5.10e-05$ & $1.75 \pm 0.12$ & $2.37 \pm 0.06$ & CP\textsuperscript{g} \\
235678745 & 2095.02 & $28.1739 \pm 0.0567$ & $28.1723 \pm 8.85e-05$ & $1.14 \pm 0.08$ & $1.45 \pm 0.52$ & CP\textsuperscript{r} \\
287080092 & 1751.01 & $37.4669 \pm 0.0006$ & $37.4685 \pm 8.20e-05$ & $1.98 \pm 0.17$ & $2.77 \pm 0.15$ & CP\textsuperscript{t} \\
229742722 & 1859.01 & $63.4833 \pm 0.0016$ & $63.4839 \pm 0.0002$ & $10.90 \pm 0.66$ & $11.15 \pm 0.51$ & CP\textsuperscript{u} \\
232608943 & 4600.01 & $82.6847 \pm 0.0014$ & $82.6869 \pm 0.0003$ & $4.53 \pm 0.34$ & $6.80 \pm 0.31$ & CP\textsuperscript{v} \\
441765914 & 2088.01 & $124.7222 \pm 0.0023$ & $124.7300 \pm 0.00065$ & $2.49 \pm 0.22$ & $3.68 \pm 0.19$ & CP\textsuperscript{g} \\
229791084 & 1864.01 & $1.6454 \pm 2.60e-05$ & $1.6453 \pm 1.00e-05$ & $16.35 \pm 0.95$ & $17.30 \pm 0.45$ & KP\textsuperscript{w} \\
165530380 & 2262.01 & $1.4937 \pm 2.35e-05$ & $1.4937 \pm 9.70e-06$ & $1.77 \pm 0.30$ & $1.62 \pm 0.12$ & PC \\
356822426 & 4174.01 & $1.5569 \pm 2.45e-05$ & $1.5570 \pm 6.00e-06$ & $8.06 \pm 0.95$ & $6.94 \pm 0.49$ & PC \\
233071926 & 1748.01 & $1.8319 \pm 2.88e-05$ & $1.8319 \pm 5.15e-06$ & $1.52 \pm 0.16$ & $1.53 \pm 0.90$ & PC \\
229605891 & 4059.01 & $2.0382 \pm 9.77e-05$ & $2.0382 \pm 6.00e-07$ & $10.75 \pm 0.71$ & $13.01$ & PC \\
229586455 & 1887.01 & $2.2002 \pm 3.44e-05$ & $2.2002 \pm 4.37e-06$ & $2.38 \pm 0.15$ & $2.50 \pm 1.03$ & PC \\
230388132 & 4057.01 & $2.4076 \pm 0.0001$ & $2.4077 \pm 4.60e-06$ & $7.39 \pm 0.60$ & $7.16$ & PC \\
219776325 & 1610.01 & $2.5288 \pm 3.95e-05$ & $2.5288 \pm 4.10e-06$ & $2.07 \pm 0.18$ & $2.30 \pm 0.13$ & PC \\
441763252 & 4468.01 & $2.7708 \pm 0.0002$ & $2.7709 \pm 1.10e-06$ & $10.43 \pm 0.73$ & $12.26$ & PC \\
219762508 & 1878.01 & $3.2351 \pm 6.36e-05$ & $3.2351 \pm 4.33e-06$ & $6.23 \pm 0.50$ & $6.97 \pm 1.16$ & PC \\
229581160 & 4452.01 & $3.4149 \pm 0.0001$ & $3.4149 \pm 8.31e-06$ & $11.34 \pm 0.79$ & $13.92 \pm 0.18$ & PC \\
232632239 & 4107.01 & $3.5325 \pm 0.0002$ & $3.5323 \pm 1.38e-05$ & $2.70 \pm 0.15$ & $3.40 \pm 0.37$ & PC \\
307956397 & 1832.01 & $4.1508 \pm 7.97e-05$ & $4.1509 \pm 4.10e-06$ & $6.57 \pm 0.49$ & $7.61 \pm 0.59$ & PC \\
359388309 & 4172.01 & $4.1915 \pm 6.56e-05$ & $4.1915 \pm 3.03e-05$ & $2.33 \pm 0.23$ & $2.33 \pm 0.18$ & PC \\
198358065 & 7281.01 & $4.3096 \pm 6.77e-05$ & $4.3096 \pm 4.30e-06$ & $3.11 \pm 0.24$ & $6.96 \pm 0.89$ & PC \\
264173803 & 4099.01 & $4.3955 \pm 6.91e-05$ & $4.3954 \pm 3.33e-05$ & $2.01 \pm 0.20$ & $2.03 \pm 0.17$ & PC \\
230377505 & 1706.01 & $4.5156 \pm 7.22e-05$ & $4.5156 \pm 1.12e-05$ & $1.90 \pm 0.18$ & $2.06 \pm 0.12$ & PC \\
233053554 & 4055.01 & $4.5677 \pm 0.0003$ & $4.5679 \pm 6.10e-06$ & $13.04 \pm 0.95$ & $14.23 \pm 0.00$ & PC \\
259172391 & 1445.02 & $4.7843 \pm 7.50e-05$ & $4.7842 \pm 3.02e-05$ & $2.01 \pm 0.18$ & $2.14 \pm 0.12$ & PC \\
229408913 & 4114.01 & $4.8043 \pm 0.0002$ & $4.8039 \pm 9.02e-05$ & $2.98 \pm 0.14$ & $4.10 \pm 0.47$ & PC \\
229781583 & 1245.01 & $4.8204 \pm 0.0001$ & $4.8205 \pm 1.22e-05$ & $2.18 \pm 0.10$ & $2.37 \pm 0.10$ & PC \\
280031353 & 2300.02 & $4.8567 \pm 7.57e-05$ & $4.8567 \pm 1.30e-05$ & $2.49 \pm 0.23$ & $3.30 \pm 0.22$ & PC \\
280035202 & 4094.01 & $4.9113 \pm 7.69e-05$ & $4.9112 \pm 5.57e-05$ & $2.90 \pm 0.28$ & $3.44 \pm 0.29$ & PC \\
233680651 & 2166.01 & $5.1354 \pm 8.06e-05$ & $5.1355 \pm 1.37e-05$ & $3.57 \pm 0.31$ & $3.34 \pm 0.17$ & PC \\
160618074 & 4119.01 & $5.2018 \pm 0.0002$ & $5.2020 \pm 1.10e-05$ & $8.12 \pm 0.65$ & $9.15 \pm 0.73$ & PC \\
441797803 & 1302.01 & $5.6667 \pm 5.02e-05$ & $5.6666 \pm 1.80e-06$ & $13.98 \pm 0.80$ & $14.86 \pm 0.62$ & PC \\
230129753 & 4459.01 & $5.7477 \pm 0.0002$ & $5.7475 \pm 6.45e-05$ & $5.34 \pm 0.25$ & $6.43 \pm 0.34$ & PC \\
288428649 & 5229.01 & $5.9184 \pm 0.0001$ & $5.9185 \pm 2.20e-05$ & $3.32 \pm 0.26$ & $3.47 \pm 0.23$ & PC \\
230075227 & 2077.01 & $6.1155 \pm 9.59e-05$ & $6.1156 \pm 2.35e-05$ & $1.38 \pm 0.12$ & $1.56 \pm 0.12$ & PC \\
289590465 & 2102.01 & $6.7061 \pm 0.0001$ & $6.7060 \pm 1.81e-05$ & $1.52 \pm 0.18$ & $2.13 \pm 0.46$ & PC \\
219866209 & 4062.01 & $6.7478 \pm 0.0003$ & $6.7474 \pm 7.20e-06$ & $11.57 \pm 0.81$ & $13.10 \pm 0.00$ & PC \\
219879302 & 1884.01 & $6.7823 \pm 0.0001$ & $6.7823 \pm 3.46e-06$ & $8.08 \pm 0.56$ & $9.05 \pm 0.54$ & PC \\
356786657 & 4040.01 & $7.1356 \pm 0.0001$ & $7.1357 \pm 0.0001$ & $2.65 \pm 0.23$ & $2.79 \pm 1.55$ & PC \\
230377505 & 1706.02 & $7.4057 \pm 0.0001$ & $7.4058 \pm 4.87e-05$ & $1.66 \pm 0.17$ & $1.58 \pm 0.14$ & PC \\
233068569 & 2181.01 & $8.3792 \pm 0.0001$ & $8.3791 \pm 1.82e-05$ & $4.30 \pm 0.31$ & $4.97 \pm 0.28$ & PC \\
462615350 & 2288.01 & $8.9171 \pm 0.0001$ & $8.9172 \pm 1.76e-05$ & $2.49 \pm 0.12$ & $2.79 \pm 0.95$ & PC \\
230017324 & 1280.01 & $9.6925 \pm 0.0002$ & $9.6926 \pm 2.08e-05$ & $2.94 \pm 0.21$ & $3.21 \pm 0.25$ & PC \\
259172391 & 1445.01 & $9.8131 \pm 0.0002$ & $9.8131 \pm 5.69e-05$ & $2.53 \pm 0.18$ & $2.67 \pm 0.15$ & PC \\
229750058 & 1825.01 & $10.1824 \pm 0.0002$ & $10.1824 \pm 5.80e-06$ & $9.68 \pm 0.68$ & $10.77 \pm 0.59$ & PC \\
284454822 & 4093.01 & $10.4720 \pm 0.0002$ & $10.4722 \pm 4.28e-05$ & $2.99 \pm 0.24$ & $3.14 \pm 0.20$ & PC \\
362103298 & 4573.01 & $10.7155 \pm 0.0002$ & $10.7156 \pm 4.88e-05$ & $1.28 \pm 0.10$ & $1.42 \pm 0.50$ & PC \\
198206613 & 1741.01 & $10.9413 \pm 0.0003$ & $10.9407 \pm 5.32e-05$ & $1.05 \pm 0.10$ & $1.36 \pm 0.14$ & PC \\
229944666 & 1464.01 & $11.3125 \pm 0.0002$ & $11.3125 \pm 1.86e-05$ & $2.51 \pm 0.18$ & $2.69 \pm 0.42$ & PC \\
232624234 & 5611.01 & $11.3625 \pm 0.0002$ & $11.3630 \pm 5.99e-05$ & $2.04 \pm 0.18$ & $2.39 \pm 0.22$ & PC \\
219875976 & 5728.01 & $11.4977 \pm 0.0002$ & $11.4975 \pm 0.0001$ & $1.16 \pm 0.09$ & $1.13 \pm 0.86$ & PC \\
219847787 & 5265.01 & $11.5913 \pm 0.0002$ & $11.5915 \pm 6.50e-05$ & $2.86 \pm 0.25$ & $3.17 \pm 0.23$ & PC \\
424391516 & 1761.01 & $11.5921 \pm 0.0002$ & $11.5923 \pm 4.04e-05$ & $2.22 \pm 0.20$ & $2.42 \pm 0.68$ & PC \\
353782445 & 1664.01 & $11.8352 \pm 0.0002$ & $11.8351 \pm 1.98e-05$ & $2.24 \pm 0.16$ & $2.52 \pm 0.14$ & PC \\
230084146 & 4109.01 & $12.3202 \pm 0.0002$ & $12.3203 \pm 3.13e-05$ & $4.15 \pm 0.16$ & $5.11 \pm 0.50$ & PC \\
233066156 & 2252.01 & $12.3797 \pm 0.0002$ & $12.3796 \pm 5.79e-05$ & $6.14 \pm 0.52$ & $8.70 \pm 0.56$ & PC \\
243335710 & 1879.01 & $13.7043 \pm 0.0003$ & $13.7046 \pm 4.98e-05$ & $13.71 \pm 0.93$ & $17.04 \pm 0.81$ & PC \\
230088370 & 1176.01 & $14.0080 \pm 0.0002$ & $14.0078 \pm 4.80e-06$ & $11.26 \pm 0.79$ & $12.76 \pm 0.54$ & PC \\
233009109 & 1737.01 & $14.4478 \pm 0.0002$ & $14.4479 \pm 7.69e-05$ & $2.70 \pm 0.21$ & $2.80 \pm 0.13$ & PC \\
280031353 & 2300.01 & $15.4434 \pm 0.0002$ & $15.4433 \pm 2.47e-05$ & $5.37 \pm 0.40$ & $5.71 \pm 0.33$ & PC \\
332477926 & 1754.01 & $16.2143 \pm 0.0003$ & $16.2143 \pm 6.85e-05$ & $2.28 \pm 0.10$ & $2.82 \pm 0.14$ & PC \\
441730540 & 4025.01 & $16.2917 \pm 0.0020$ & $16.2841 \pm 0.0001$ & $7.59 \pm 0.66$ & $7.99 \pm 0.00$ & PC \\
441804533 & 5711.01 & $16.3038 \pm 0.0035$ & $16.3037 \pm 5.03e-05$ & $2.10 \pm 0.10$ & $2.25 \pm 0.83$ & PC \\
233719215 & 4054.01 & $17.3576 \pm 0.0013$ & $17.3567 \pm 9.45e-05$ & $7.57 \pm 0.60$ & $9.71 \pm 0.23$ & PC \\
284454822 & 4093.02 & $17.7044 \pm 0.0004$ & $17.7046 \pm 0.0004$ & $2.83 \pm 0.25$ & $2.86 \pm 1.11$ & PC \\
233633993 & 2083.01 & $17.9700 \pm 0.0004$ & $17.9696 \pm 8.35e-05$ & $2.25 \pm 0.18$ & $3.06 \pm 0.36$ & PC \\
356016119 & 2094.01 & $18.7932 \pm 0.0003$ & $18.7932 \pm 3.84e-05$ & $1.43 \pm 0.10$ & $1.72 \pm 0.69$ & PC \\
459969957 & 1274.01 & $19.3205 \pm 0.0003$ & $19.3203 \pm 1.26e-05$ & $8.77 \pm 0.63$ & $9.39 \pm 0.69$ & PC \\
230387153 & 2086.02 & $20.2622 \pm 0.0003$ & $20.2617 \pm 9.31e-05$ & $2.24 \pm 0.24$ & $2.42 \pm 0.73$ & PC \\
420112217 & 4163.01 & $21.5622 \pm 0.0003$ & $21.5615 \pm 0.0001$ & $8.10 \pm 0.63$ & $7.76 \pm 0.43$ & PC \\
459970307 & 1154.01 & $21.7452 \pm 0.0003$ & $21.7455 \pm 4.96e-05$ & $2.17 \pm 0.16$ & $2.10 \pm 0.11$ & PC \\
229786610 & 4113.01 & $22.1582 \pm 0.0004$ & $22.1575 \pm 0.0001$ & $2.21 \pm 0.18$ & $2.17 \pm 0.17$ & PC \\
383645563 & 5225.01 & $22.9398 \pm 0.0043$ & $22.9367 \pm 0.0001$ & $10.23 \pm 1.43$ & $8.88 \pm 0.76$ & PC \\
198189972 & 4117.01 & $25.0938 \pm 0.0004$ & $25.0930 \pm 0.0003$ & $8.60 \pm 0.69$ & $10.31 \pm 0.64$ & PC \\
259100469 & 5233.01 & $25.7627 \pm 0.0004$ & $25.7633 \pm 6.82e-05$ & $5.14 \pm 0.41$ & $5.77 \pm 0.34$ & PC \\
233688779 & 2254.01 & $26.4583 \pm 0.0004$ & $26.4580 \pm 0.0001$ & $2.53 \pm 0.21$ & $2.91 \pm 0.26$ & PC \\
232624234 & 5611.02 & $26.7173 \pm 0.0004$ & $26.7160 \pm 0.0003$ & $2.30 \pm 0.22$ & $2.74 \pm 1.69$ & PC \\
224596152 & 1734.01 & $28.8741 \pm 0.0005$ & $28.8745 \pm 0.0001$ & $2.30 \pm 0.19$ & $2.71 \pm 0.20$ & PC \\
259274960 & 4049.01 & $31.4038 \pm 0.0024$ & $31.4001 \pm 0.0003$ & $10.00 \pm 0.86$ & $10.25 \pm 0.45$ & PC \\
320324079 & 5202.01 & $34.6154 \pm 0.0006$ & $34.6142 \pm 0.0003$ & $6.22 \pm 0.58$ & $7.70 \pm 1.73$ & PC \\
230387153 & 2086.01 & $41.7547 \pm 0.0007$ & $41.7540 \pm 0.0004$ & $2.55 \pm 0.22$ & $2.75 \pm 0.20$ & PC \\
219803922 & 7025.01 & $66.9945 \pm 0.0016$ & $66.9954 \pm 0.0001$ & $14.32 \pm 1.25$ & $11.72 \pm 0.65$ & PC \\

\midrule
\multicolumn{7}{p{0.95\textwidth}}{\footnotesize
\textsuperscript{a} \cite{rubenzahl_tess-keck_2024};
\textsuperscript{b} \cite{pelaez-torres_gem_2026};
\textsuperscript{c} \cite{brady_early_2024};
\textsuperscript{d} \cite{lafarga_automatic_2026};
\textsuperscript{e} \cite{ehrhardt_confirmation_2024};
\textsuperscript{f} \cite{bakos_hat-p-58b-hat-p-64b_2021};
\textsuperscript{g} \cite{polanski_tess-keck_2024};
\textsuperscript{h} \cite{yalcinkaya_toi-1743_2025};
\textsuperscript{i} \cite{stalport_tess_2025};
\textsuperscript{j} \cite{fukui_toi-1749_2021};
\textsuperscript{k} \cite{ghachoui_tess_2023};
\textsuperscript{l} \cite{persson_toi-1438_2025};
\textsuperscript{m} \cite{gomez_barrientos_validation_2025};
\textsuperscript{n} \cite{barkaoui_toi-2084_2023};
\textsuperscript{o} \cite{crossfield_orcas_2025};
\textsuperscript{p} \cite{macdougall_tess-keck_2021};
\textsuperscript{q} \cite{cadieux_toi-1452_2022};
\textsuperscript{r} \cite{murgas_two_2023};
\textsuperscript{s} \cite{pidhorodetska_tess-keck_2024};
\textsuperscript{t} \cite{desai_tess-keck_2024}
\textsuperscript{u} \cite{dong_toi-1859b_2023};
\textsuperscript{v} \cite{mireles_toi-4600_2023};
\textsuperscript{w} \cite{alsubai_qatar_2019}.}

\end{longtable}
\end{center}

\begin{center}
\scriptsize
\setlength{\tabcolsep}{4pt}
\renewcommand{\arraystretch}{1.15}

\begin{longtable}{
@{}
r
r
c
c
>{\centering\arraybackslash}p{0.035\linewidth}
@{\hspace{28pt}}
>{\centering\arraybackslash}p{0.38\linewidth}
@{}
}

\caption{Unrecovered TOIs and their ExoFOP orbital periods and planet radii.}
\label{tab:unrecovered_TOIs}\\

\toprule
TIC ID & TOI ID &
$P$ (TOI; days) &
$R_p$ (TOI; $R_\oplus$) &
TFOP Dis. & Reason missed \\
\midrule
\endfirsthead

\toprule
TIC ID & TOI ID &
$P$ (TOI; days) &
$R_p$ (TOI; $R_\oplus$) &
TFOP Dis. & Reason missed \\
\midrule
\endhead

\midrule
\multicolumn{6}{r}{\footnotesize Continued on next page}\\
\endfoot

\bottomrule
\endlastfoot

\multicolumn{6}{c}{\textbf{Unrecovered TOIs}}\\
\midrule

235683377 & 1442.01 & $0.4091 \pm 3.00e-07$ & $1.17 \pm 0.06$ & CP\textsuperscript{x} & BLS recovered an incorrect period; EM2 and 2-min tests did not resolve the mismatch. \\
219852882 & 1346.01 & $1.7623 \pm 4.70e-06$ & $2.39 \pm 0.16$ & CP\textsuperscript{m} & BLS recovered an incorrect period; EM2 and 2-min tests did not resolve the mismatch. \\
441739020 & 1670.02 & $10.9846 \pm 5.10e-04$ & $2.06 \pm 0.19$ & CP\textsuperscript{y} & Detected by BLS, but failed vetting; recovered with a longer detrending window. \\
287139872 & 1752.02 & $32.7144 \pm 0.0004$ & $2.29 \pm 0.14$ & CP\textsuperscript{b} & BLS SNR below adopted threshold. \\
441739020 & 1670.01 & $40.7498 \pm 2.10e-04$ & $11.06 \pm 0.28$ & CP\textsuperscript{y} & Detected by BLS, but failed vetting due to odd-even and weak/inconsistent transit tests. \\
232608943 & 4600.02 & $482.8191 \pm 0.0018$ & $9.42 \pm 0.42$ & CP\textsuperscript{v} & Detected by BLS, but failed vetting with only two clear transits; literature confirmation used 2-min and ground-based follow-up. \\

233602827 & 1749.03 & $2.3889 \pm 9.96e-06$ & $1.34 \pm 2.03$ & KP\textsuperscript{j} & BLS SNR below adopted threshold, although follow-up observations exist. \\

198212955 & 1242.01 & $0.3815 \pm 6.00e-07$ & $2.06 \pm 0.26$ & PC & BLS recovered an incorrect period; EM2 and 2-min tests did not resolve the mismatch. \\
259233660 & 6000.01 & $0.4490 \pm 1.11e-05$ & $1.02 \pm 1.28$ & PC & BLS recovered an incorrect period; EM2 and 2-min tests did not resolve the mismatch. \\
219742885 & 4580.01 & $0.9168 \pm 1.23e-05$ & $0.61 \pm 1.02$ & PC & Detected by BLS, but failed vetting due to weak and inconsistent transit SNR. \\
356871098 & 5983.01 & $1.0272 \pm 7.45e-06$ & $1.38 \pm 0.17$ & PC & Detected by BLS, but failed vetting due to inconsistent transit SNR. \\
233211762 & 1252.01 & $1.1220 \pm 1.36e-06$ & $4.00 \pm 0.67$ & PC & Detected by BLS, but failed multiple vetting tests and was flagged as off-target. \\
229938290 & 1783.01 & $1.4199 \pm 0.0002$ & $2.28 \pm 1.69$ & PC & BLS recovered an incorrect period; EM2 and 2-min tests did not resolve the mismatch. \\
237101326 & 4051.01 & $1.5374 \pm 3.00e-06$ & $5.76$ & PC & BLS recovered an incorrect period; EM2 and 2-min tests did not resolve the mismatch. \\
230087765 & 5740.01 & $1.6894 \pm 1.53e-05$ & $1.38 \pm 2.16$ & PC & BLS SNR below adopted threshold. \\
356916207 & 6044.01 & $2.2539 \pm 0.0006$ & $2.20 \pm 7.07$ & PC & BLS SNR below adopted threshold. \\
233071822 & 1747.01 & $2.7547 \pm 3.90e-06$ & $2.17 \pm 0.48$ & PC & Detected by BLS, but failed vetting due to inconsistent transit SNR. \\
230375747 & 1870.01 & $2.7634 \pm 1.33e-06$ & $16.41 \pm 1.11$ & PC & Detected by BLS, but failed multiple vetting tests. \\
441798995 & 2269.02 & $2.8411 \pm 1.33e-05$ & $1.42 \pm 0.12$ & PC & BLS recovered an incorrect period; EM2 and 2-min tests did not resolve the mismatch. \\
199712572 & 6082.01 & $3.5942 \pm 1.59e-05$ & $1.31 \pm 0.10$ & PC & BLS recovered an incorrect period; EM2 and 2-min tests did not resolve the mismatch. \\
441739871 & 1763.01 & $3.7979 \pm 8.78e-06$ & $1.93 \pm 0.84$ & PC & BLS SNR below adopted threshold. \\
198187049 & 2069.01 & $5.9214 \pm 1.43e-05$ & $1.05 \pm 0.32$ & PC & BLS recovered an incorrect period; EM2 and 2-min tests did not resolve the mismatch. \\
233078665 & 1876.01 & $6.4828 \pm 6.70e-06$ & $11.49 \pm 0.80$ & PC & Detected by BLS, but failed vetting due to a significant secondary. \\
441763252 & 4468.02 & $7.0140 \pm 0.0001$ & $3.65 \pm 0.45$ & PC & BLS SNR below adopted threshold. \\
287152495 & 6253.01 & $9.2347 \pm 1.88e-05$ & $1.39 \pm 3.46$ & PC & BLS recovered an incorrect period; EM2 and 2-min tests did not resolve the mismatch. \\
389968080 & 4031.01 & $10.0463 \pm 4.11e-05$ & $16.45 \pm 3.19$ & PC & Detected by BLS, but failed vetting due to non-unique events and inconsistent SNR. \\
441798995 & 2269.03 & $10.4029 \pm 3.17e-05$ & $1.85 \pm 0.27$ & PC & BLS SNR below adopted threshold. \\
233047097 & 1875.01 & $11.3092 \pm 1.42e-05$ & $15.44 \pm 0.96$ & PC & Detected by BLS, but failed vetting because the inferred radius was too large. \\
230006888 & 4111.01 & $12.2788 \pm 5.95e-05$ & $9.65 \pm 0.59$ & PC & Detected when using 10-min cadence rather than the binned 30-min light curve. \\
237211410 & 2282.01 & $12.5521 \pm 4.90e-05$ & $2.82 \pm 0.25$ & PC & Detected by BLS, but failed pixel vetting and FA tests; light curve is highly scattered. \\
229618478 & 5737.01 & $19.4481 \pm 8.78e-05$ & $2.30 \pm 0.74$ & PC & BLS recovered an incorrect period; EM2 and 2-min tests did not resolve the mismatch. \\
233500935 & 2297.01 & $22.0488 \pm 0.0002$ & $1.95 \pm 0.99$ & PC & BLS SNR below adopted threshold. \\
353807936 & 5650.01 & $34.2826 \pm 0.0002$ & $5.85 \pm 0.34$ & PC & Detected by BLS, but failed vetting due to a poor transit model fit. \\
284900292 & 5987.01 & $48.7131 \pm 0.0005$ & $2.49 \pm 1.02$ & PC & BLS recovered an incorrect period; EM2 and 2-min tests did not resolve the mismatch. \\
376867898 & 7064.01 & $61.5556 \pm 0.0007$ & $2.18 \pm 0.15$ & PC & BLS recovered an incorrect period; EM2 and 2-min tests did not resolve the mismatch. \\
198456933 & 2276.01 & $76.8664 \pm 0.0007$ & $1.92 \pm 0.17$ & PC & BLS recovered an incorrect period; EM2 and 2-min tests did not resolve the mismatch. \\
441739020 & 1670.03 & $123.0586 \pm 0.0011$ & $2.95 \pm 0.74$ & PC & BLS recovered an incorrect period; EM2 and 2-min tests did not resolve the mismatch. \\
165554103 & 2296.01 & $139.2632 \pm 0.0052$ & $1.84 \pm 0.80$ & PC & BLS SNR below adopted threshold. \\
219778329 & 2091.01 & $177.2189 \pm 0.0078$ & $2.03 \pm 1.13$ & PC & BLS recovered an incorrect period; EM2 and 2-min tests did not resolve the mismatch. \\
237201858 & 2286.01 & $179.4032 \pm 0.0014$ & $3.75 \pm 0.68$ & PC & BLS SNR below adopted threshold. \\
264171144 & 2281.01 & $210.6684 \pm 0.0021$ & $2.90 \pm 0.96$ & PC & BLS recovered an incorrect period; EM2 and 2-min tests did not resolve the mismatch. \\
233721037 & 2090.01 & $230.9221 \pm 0.0153$ & $3.68 \pm 0.95$ & PC & BLS SNR below adopted threshold. \\
376847633 & 5975.01 & $373.0382 \pm 0.0010$ & $8.45 \pm 0.47$ & PC & EM2 refined the period/epoch, but the signal still failed vetting; only two clean PM--EM1 transits. \\
424388628 & 6075.01 & $832.9236 \pm 0.0024$ & $1.74 \pm 0.84$ & PC & Only two transits, with one in Sector 75 outside the PM--EM1 search baseline. \\

\midrule
\multicolumn{6}{p{0.95\textwidth}}{\footnotesize
\textsuperscript{b} \cite{pelaez-torres_gem_2026};
\textsuperscript{j} \cite{fukui_toi-1749_2021};
\textsuperscript{m} \cite{gomez_barrientos_validation_2025};
\textsuperscript{v} \cite{mireles_toi-4600_2023};
\textsuperscript{x} \cite{giacalone_validation_2022};
\textsuperscript{y} \cite{tran_toi-1670_2022}.}

\end{longtable}
\end{center}

\clearpage
%%%%%%%%%%%%%%%%%%%%%%%%%%%%%%%%%%%%%%%%%%%%%%%%%%%%%%%
%%%%%%%%%%%%%%%%%%% References %%%%%%%%%%%%%%%%%%%%%%%
%%%%%%%%%%%%%%%%%%%%%%%%%%%%%%%%%%%%%%%%%%%%%%%%%%%%%%%
% \bibliographystyle{apj}
\bibliographystyle{aasjournalv7}
\bibliography{main, ref}

\end{document}